\documentclass[11pt,a4paper]{article}
\title{Differential Dynamic Microscopy can be applied to Differential Interference Contrast images despite shadowing effects}
\usepackage{amsmath}
\usepackage{mathtools}
\usepackage{graphicx}
\usepackage{amsthm}
\usepackage{slashed}
\usepackage{lineno}
\usepackage[section]{placeins}
\usepackage{latexsym}
\usepackage{hyperref}
\usepackage{amssymb}
\usepackage{gensymb}

\newcommand{\bb}{\begin{equation}}
\newcommand{\ee}{\end{equation}}

\usepackage{multirow}
\usepackage{ctable}
\usepackage{bm}
\usepackage{enumerate}
\usepackage[square,sort,comma,compress,numbers]{natbib}

\newcommand{\D}[2]{\frac{\partial #1}{\partial #2}}

\usepackage{framed}

\usepackage{url}
\usepackage{soul}

\newcommand{\mum}{\mu \text{m}^{-1}}
\newcommand{\diffunit}{\mu \text{m}^{2}/\text{s}}

\usepackage{setspace}
\usepackage[geometry]{ifsym}
\usepackage{float}
\usepackage{subcaption}
\usepackage{placeins}
\usepackage{textcomp}
\usepackage[parfill]{parskip}
\usepackage[left=2cm,right=2cm,top=2cm,bottom=3cm]{geometry}
\providecommand{\keywords}[1]
{
  \small	
  \textbf{\textit{Keywords---}} #1
}
\usepackage{url}
\usepackage{cleveref}
	\author{%%%% Author details
	Timothy Ostler$^{1}$, Wolfgang Langbein$^{2}$, Paola Borri$^{3}$, Emily Lewis$^{3}$, \\ Yisu Wang$^{3}$, Karl Swann $^{3}$, Thomas E. Woolley$^{1}$, Katerina Kaouri$^{1}$}
 \date{%
 $^{1}$Cardiff School of Mathematics, Cardiff University, Cardiff, United Kingdom\\
 $^{2}$Cardiff School of Physics and Astronomy, Cardiff University, Cardiff, United Kingdom\\
 $^{3}$Cardiff School of Biosciences, Cardiff University, Cardiff, United Kingdom} 
\usepackage{graphicx}
\graphicspath{{Images/}}

\newcommand{\Iout}{\ensuremath{I_{\rm t}}}
\newcommand{\Iex}{\ensuremath{I_{\rm e}}}
\newcommand{\br}{\ensuremath{\mathbf{r}}}
\newcommand{\bs}{\ensuremath{\mathbf{s}}}
\newcommand{\Na}{\ensuremath{N_{\rm a}}} 
\begin{document}
	\maketitle
	\delimitershortfall=-1pt
%TC:endignore	
\begin{abstract}
   During In Vitro Fertilisation (IVF), combining time-lapse contrast microscopy with suitable image processing techniques could facilitate non-invasive classification of oocyte health, driving oocyte selection to maximise success rates. One image processing technique, Differential Dynamic Microscopy (DDM), has been used in a variety of microscopy settings including bright field, dark field and fluorescence. However, in some applications, effects stemming from the choice of microscopy may invalidate underlying assumptions of DDM. Here, we study the DDM analysis of differential interference contrast (DIC) microscopy movies. DIC exhibits a characteristic shadowing effect that gives the illusion of a 3D appearance, which we show causes deformation of the output DDM matrix. We present a mathematical description of this deformation, and conclude that, when studying isotropic motion, no account of the DIC shadow needs to be considered during DDM analysis. We validate our conclusions with simulated particle data, and with DIC images of colloidal dispersions. Although this conclusion does not generally extend to anisotropic motion, we demonstrate that for directed advection and diffusion behaviour, parameter fitting invariance still holds. These results validate the current practice of applying DDM to DIC, and are a foundation for further exploration of DDM in other phase-contrast image datasets.
\end{abstract}
\section{Introduction}
\label{Section: introduction}

Quantifiable assessment of the rate of movement within image data is required in a wide variety of medical diagnostic settings. For example, characterising spermatozoa motility is crucial for fertility assessment of both farm animals and humans \cite{Jepson2019}, and there are established links between the rate of so-called basal speed, or `active diffusion', in the cytoplasm of mouse zygotes and their potential for embryo development \cite{Ajduk_2011}. Although important, the usefulness of such data depends on the availability of suitable techniques that can accurately extract quantifiable movement data from images. 

Differential Dynamic Microscopy (DDM) is one such technique, designed to interpret and analyse time lapse microscopy images to extract statistical motility parameters \cite{Cerbino2008}. DDM has been applied to a number of problems, including characterising the diffusion rate of Brownian motion in colloidal dispersions \cite{Cerbino2008}, describing the swimming speed distribution of bacteria \cite{Wilson_2011} and spermatozoa \cite{Jepson2019}, and characterising active diffusive processes in Drosophila oocytes \cite{Drechsler_2017}. The applications of DDM cover many types of microscopy, including bright field \cite{Cerbino2008}, fluorescence \cite{He2012}, confocal fluorescence \cite{Lu_2012}, dark-field \cite{Bayles_2016} and phase contrast \cite{Wilson_2011, Martinez_2012}. 

\begin{figure}[H]
	\centering
	\includegraphics[width=0.6\linewidth]{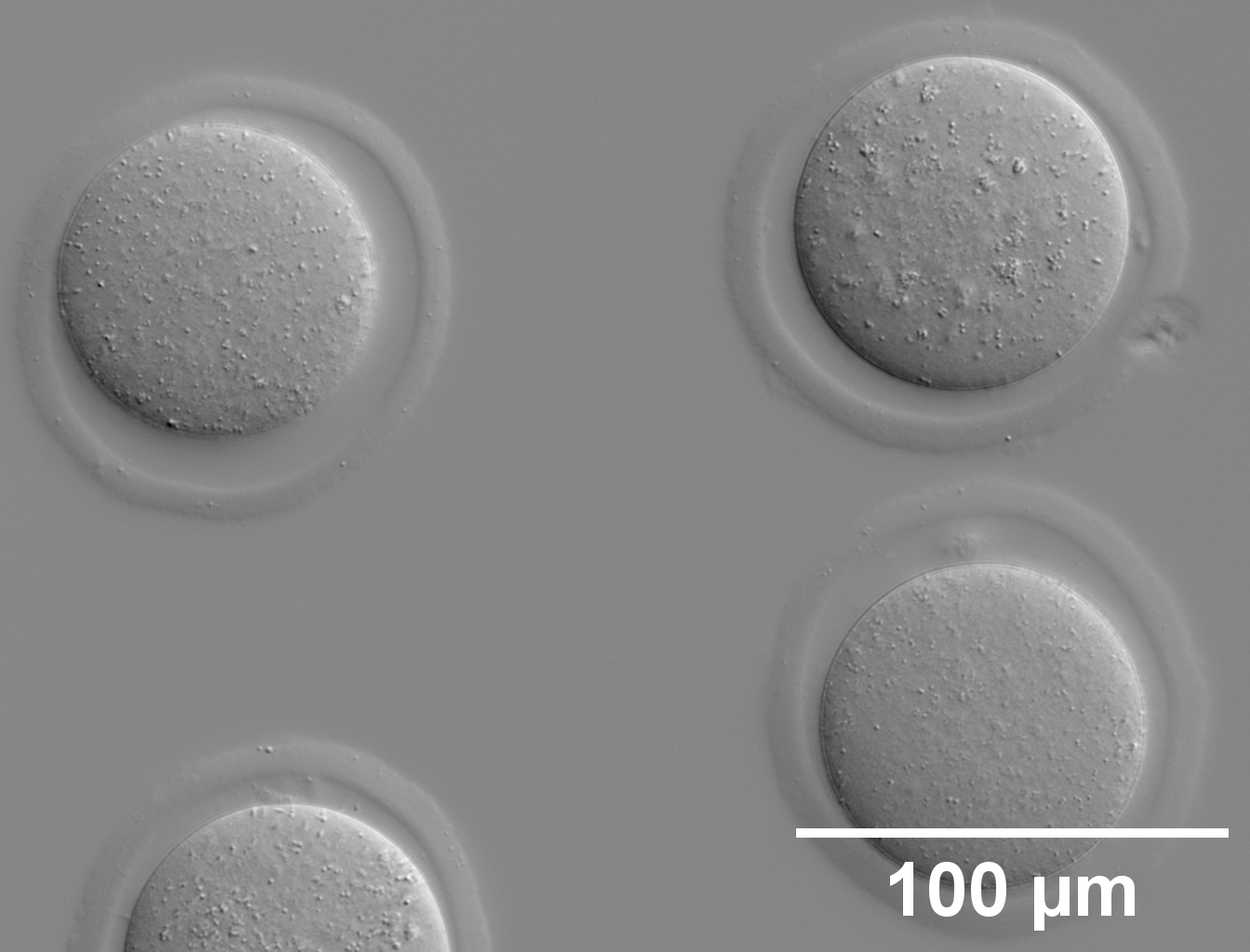}
	\caption{Several mouse oocytes imaged with DIC microscopy, as described in Section \ref{sec:Setup}. A visual shadow can be seen, as if a light source were supplied at the top right of the image. Image by Wang, Borri and Langbein.}
	\label{fig: MouseOocyteDIC}
\end{figure}

The specific phase contrast application we consider here is characterising the rate of cytoplasmic flow in Drosophila oocytes \cite{Drechsler_2017}, which could pave the way for non-invasive techniques to assess IVF success. This application uses differential interference contrast (DIC) microscopy, which utilises dual beam interference to convert phase gradients into light and dark intensities \cite{Murphy_2012,Hoffman_1975}, and can generate images with very high resolution and contrast \cite{Allen1969}, even at very large magnification \cite{Hamilton_2022}. A characteristic feature of DIC is that the images appear `shadow-cast' \cite{Murphy_2012} as demonstrated in Figure \ref{fig: MouseOocyteDIC}, depicting a DIC image of a mouse oocyte.

DIC is bright field transmission modulated by the sample phase gradients along a given direction, the shear. This contrast is similar to the relief-type contrast of the brightness of a surface profile with a height representing the phase, obliquely illuminated along the shear, providing highlights and shadows, and thus can be understood by human observers as such a surface profile, which they are trained to see during their development \cite{Murphy_2012}. The anisotropy introduced by the shear raises the question of how to interpret the output statistics from image processing techniques such as DDM in terms of the underlying object motion. For example, in dark-field imaging, DDM is only robust under the condition that gradients in the background illumination are small relative to the length scale of imaged motion \cite{Bayles_2016}. Although phase contrast microscopy has been used for DDM in various studies \cite{Wilson_2011,Martinez_2012,Reufer_2012,Drechsler_2017}, the authors of this paper are not aware of investigations into how the shade-off and halo artifacts in traditional phase contrast microscopy using annular illumination and a phase ring \cite{Murphy_2012} affects the output statistics from DDM.

This work evaluates the use of DDM on DIC movies and demonstrates the effects of the changed contrast on the DDM algorithm. Importantly, we show that current DDM analysis approaches will yield equivalent summary statistics under certain conditions. First, we give an overview of DIC microscopy in Section \ref{Section: DIC}, including methods for acquisition of images used in this work. We then give a brief overview of the DDM technique in Section \ref{section: DDM}, and establish a mathematical formulation for the DDM matrix generated by a DIC image stack. From this formulation, in Section \ref{section: isotropic} we immediately prove that in the case of isotropic motion, the DDM analysis is invariant under the DIC shadowing, up to a scaling. We further extend this conclusion in Section \ref{Section: AdvectionDIC} to include a system with advection and diffusion, which constitutes a specific case of anisotropic motion.

%(This justifies the application of DDM without adjustment to the case of Drosophila oocytes \cite{Drechsler_2017}.)

\section{DIC Microscopy}\label{Section: DIC}

\begin{figure}
	\centering
	\includegraphics[width=.7\linewidth]{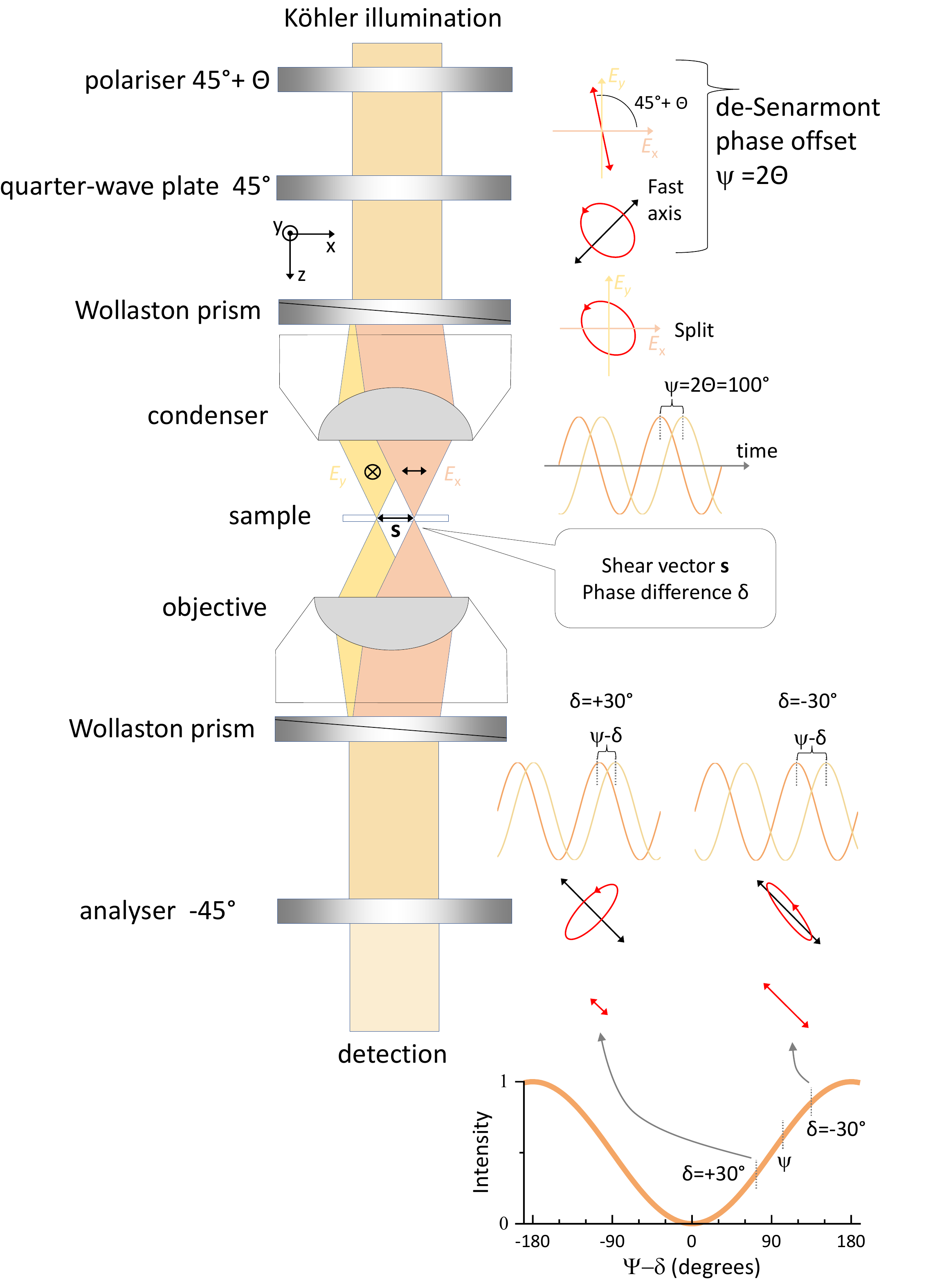}
	\caption{A sketch of the principle of operation of a DIC microscope with de-S\'enarmont compensation. After adjusting a transmission microscope for K\"ohler illumination, a linear polariser at $45\degree+\theta$ to the sample $x$ (horizontal) and $y$ (out of plane) directions creates a linear polarisation (see electric field sketches on the right), which is converted by a quarter wave plate to a relative phase shift of $2\theta$ of the $x$ and $y$ polarised field components $E_x$ and $E_y$ of equal amplitude (see sketch of temporal field oscillations on the right). A Wollaston prism splits the two fields, creating two focal points separated by the shear \bs. The transmitted fields, which experienced a relative phase shift $\delta(\br)$ as function of sample position $\br=(x,y)$, are collimated by the objective, recombined in direction by a second Wollaston prism, and projected by the analyser of $-45\degree$ orientation, providing an intensity proportional to $1-\cos(\psi-\delta(\br))$. Using $\psi=90\degree$, a nearly linear dependence of the intensity on the phase shift $\delta(\br)$ is obtained, leading to the relief-type contrast.}
	\label{Fig: DIC_Sketch}
\end{figure}

\subsection{DIC Principle}
\label{sec:dic}
A DIC image is formed by pairs of closely spaced polarised beams \cite{Murphy_2012}, with their separation denoted as the `shear', which is typically similar to optical resolution in the focal plane. The beams are then recombined and analysed, converting a mutual phase difference into an intensity difference in the resulting image.

A sketch of a DIC microscope with de-S\'enarmont compensation and its workings is given in Figure \ref{Fig: DIC_Sketch}. Starting from a bright-field transmission microscope in K\"ohler illumination, the illumination is linearly polarised at $45\degree+\theta$, and then transmitted through a quarter-wave plate with the fast axis at 45\degree, resulting in electric field components $E_x$ and $E_y$ in $x$ and $y$ direction, respectively, of equal amplitude and a relative phase shift of $\psi=2\theta$. A Wollaston prism splits the propagation direction of $E_x$ and $E_y$ in the condenser back focal plane, resulting in a relative displacement of the corresponding focal points at the sample by the shear \bs\ along the $x$ direction. The fields are probing the sample properties accordingly at two positions displaced by \bs, and acquiring a spatially dependent phase difference $\delta(\br)$, with the in-plane sample position $\br=(x,y)$. After transmission through the sample, the two fields are recombined in direction at the back focal plane of the objective by a Wollaston prism matched to the first one (a Nomarski prism can be used instead, which allows to have the effective recombination position displaced from the prism and thus to position the prism at a accessible place in the beam path after the objective). A second polariser, called analyser, orientated at -45\degree, orthogonal to the first one, is projecting the recombined fields along its axis to provide an intensity interference. 
The resulting transmitted intensity \Iout\ has a dependence on the phase shift $\delta(\br)$ given by
\begin{equation}
\label{eq:Iout} \Iout(\br,\psi)=\frac{\Iex}{2} \left[1-\cos\left(
\psi -\delta(\br)\right)\right]\,,
\end{equation}
with the excitation intensity \Iex, the position in the sample plane
\br, the phase offset $\psi$, and the difference $\delta(\br)$ of the
optical phase shift $\phi$ for the two beams that pass through
the sample in two adjacent points separated by the shear vector
\bs.

Adjusting the phase offset $\psi$, the contrast can be changed from a dark-field type at $\psi=0$, where the intensity is proportional to $\delta^2(\br)$, to a bright field image modulated by a term linear with the phase shift, with the largest linear range for $\psi=90\degree$, for a polariser angle of $\theta=45\degree$.    

More details on the theory of DIC can be found in Ref.\,\cite{HolmesAO87} for two-dimensional samples, in Ref.\,\cite{Preza1999} for three-dimensional samples, and a fully vectorial treatment of the fields is given in Ref.\,\cite{MunroOE05}.

\subsection{Samples}
\label{sec:Samples}
 As test standard for the DDM analysis, polystyrene (PS) beads, having a nominal radius of 100\,nm with less than 3\% coefficient of variance (cv), were purchased (Alpha Nanotech Colloidal PS Beads NP-PA07CPSX78). These PS beads were dispersed in 30\% v/v glycerol/water mixtures to a particle concentration of 0.1 mg/mL, 13\,\textmu l were pipetted into a 0.12\,mm high and 13\,mm diameter chamber made of a Grace Bio-Labs (Bend, US) SecureSeal imaging spacer on a $(76 \times 26 \times 1)$\,mm$^3$ microscope slide (Menzel Gl\"aser). The chamber was then sealed with a ($24\times24$)\,mm$^2$ \#1.5 coverslip (Menzel Gl\"aser), and stored in a 100\% humidity environment at 7\degree C until use. 

8 week old CD1 mice were intraperitoneally injected with 5IU pregnant mare’s serum gonadotrophin (PMSG) to induce ovarian follicle development. They were again injected with 10IU human chorionic gonadotrophin (hCG) approximately 48hrs later to induce ovulation. Ovulated mature (MII) eggs were collected from oviducts approximately 15hrs later. All animals were handled according to UK Home Office regulations, and procedures were carried out under a UK Home Office Project License with the approval of Cardiff University Animal Ethics Committee. Live imaging experiments were carried out with mature eggs cultured in HKSOM as described previously \cite{Wang_2022}. All drops were covered with mineral oil (embryo tested, Sigma) to prevent evaporation.
\subsection{Optical Setup}
\label{sec:Setup}
DIC images were obtained using an inverted Nikon Ti-U microscope as in \cite{Regan_2019,Hamilton_2022}. Samples were illuminated using a 100\,W halogen lamp (Nikon V2-A LL 100\,W) followed by a Schott BG40 filter to remove wavelengths above 650\,nm (for which the DIC polarisers are not suited) and a Nikon green interference filter (Nikon GIF), to define the wavelength range centred at 550\,nm and having a full-width at half maximum (FWHM) of 53\,nm. This illumination was then passed through a de-S\'enarmont compensator (a rotatable linear polariser and quarter-wave plate, Nikon T-P2 DIC Polariser HT MEN51941) and a Nomarski prism (Nikon N2 DIC module MEH52400 or MEH52500) and focused onto the sample by a condenser (MEL56100) of 0.72\,numerical aperture (NA). The shear of the Nikon N2 DIC was measured to be $(238 \pm 10)$\,nm. The objective used was a $20\times$ 0.75\,NA planapochromat (MRD00205). After the objectives, light passes through a suited Nomarski prism (DIC slider MBH76220) and a linear polariser (Nikon Ti-A-E DIC analyser block MEN 51980). Images were detected by a Hamamatsu Orca 285 CCD camera (18,000 electrons full well capacity, 7 electrons read noise, and 4.6 electrons per count, 12 bit digitiser, $1344 \times 1024$ pixels, pixel size 6.45\,$\mu$m, 192 counts offset).
The microscope was adjusted for K\"ohler illumination, and the field aperture was set to be slightly larger than the imaged sample region.

For the images of colloids, a sequence of $\Na$ frames (up to 256) were acquired, with 120\,ms exposure time per frame (given by highest stably achievable frame rate), with the lamp intensity adjusted to provide a mean intensity of about 3000 counts (13000 photoelectrons) per pixel. Data for de-S\'enarmont polariser angles $\pm\theta$ as well as zero were taken to enable qDIC analysis, for $\theta$ of 15, 30, and 45 degrees.  The rotation of the de-S\'enarmont polariser was motorized (by a home-built modification) to improve positioning speed and reproducibility. DIC images of the colloids were acquired with a 20 x objective and 1.5 x tubular lens at a polarisation angle of 15°. For DM analysis, a total of 7680 frames were taken at +15° and an average of 256 frames at -15° were used to post process the data.

For the mouse oocyte in Figure \ref{fig: MouseOocyteDIC}, DIC images were acquired with 20x objective with 1.5 tubular lens, by a monochrome ORCA285 Hamamatsu CCD camera with 120 ms frame exposure time and a 12.5° polarisation angle in the de-Senarmont DIC illuminator (yielding a 25° phase offset). A motorized stage and z-drive (Prior ProScan III) enabled lateral xy sample movement and axial z focus movement for focusing. For DM analysis, polarisation angle was changed to the opposite side every 30s, during which 256 frames of DIC images were acquired.

\section{Differential Dynamic Microscopy (DDM) applied to DIC images}
\label{section: DDM}
\subsection{The DDM algorithm}
Comprehensive descriptions of the DDM technique are available in literature \cite{Giavazzi_2014}, alongside MATLAB/Python codes and datasets \cite{Germain_2016}. Here, we provide a brief overview of the DDM algorithm.

Consider an image stack $I(\bm{r},t)$ consisting of $N$ digital images that are $L \times W$ pixels long and wide, respectively, taken at regularly spaced times $t\in\{0$, $\Delta t_{\text{min}}$, $2 \Delta t_{\text{min}}$ ... $(N-1) \Delta t_{\text{min}}\} = T$, where $t_{\text{min}}$ is the time between consecutive frames. Here, $\bm{r}$ denotes the Cartesian location of each pixel. Define the time between any two given frames as $\Delta t$, if $t + \Delta t \in T$. Then define the image difference 
\[\Delta I(\bm{r},t,\Delta t) = I(\bm{r},t+\Delta t) - I(\bm{r},t).\]
Denoting the spatial Fourier transform by $\tilde{\cdot}$, the spatial frequency vector by $\bm{q}=(q_x,q_y)$, and the time average by $\langle \cdot \rangle_t$, we define the DDM tensor as
\begin{align}
	\mathcal{D}(\bm{q},\Delta t) = \left\langle|\Delta \tilde{I}|^2\right\rangle_t = \frac{1}{N-n} \sum_{k=1}^{N-n} \left|\Delta \tilde{I}(\bm{q},k \Delta t_m, \Delta t)\right|^2, \label{eq: DDM}
\end{align}
where $n$ is the number of frames elapsed during the time lag $\Delta t$. Denoting the complex conjugate by $\cdot^*$ and commutative element-wise matrix multiplication (Hadamard product) by $\odot$, the DDM tensor can be expanded and written as \cite{Cerbino2008, Germain_2016}
\begin{align}
	\mathcal{D}(\bm{q},\Delta t) &= 2 \left\langle \left|\tilde{I}\left(\bm{q},t \right)\right|^2 \right\rangle_t \odot \left(1 -  \frac{\Re \left(\left\langle \tilde{I}^*\left(\bm{q},t+\Delta t\right) \odot \tilde{I} \left(\bm{q},t \right) \right\rangle_t \right)}{ \left\langle |\tilde{I}\left(\bm{q},t \right)|^2 \right\rangle_t} \right)  \label{eq: DDM_Decomp}.
\end{align}
Unless otherwise stated, throughout this work matrix multiplication will be element wise, so we drop the explicit $\odot$ notation. The term 
\begin{align}
	\frac{\Re(\langle \tilde{I}^* (\bm{q},t+\Delta t)  \tilde{I}(\bm{q},t) \rangle_t)}{ \langle |\tilde{I}|^2 \rangle_t} = f(\bm{q},\Delta t), \label{def: ISF}
\end{align}
is the real part of the Intermediate Scattering Function (ISF) \cite{Germain_2016,Giavazzi_2014}, which is equal to the real part of the Fourier transform of the van Hove function $G(\bm{r},t)$, which counts the expected number of particles $j$ found at the location $\bm{r}$ at time $t$ given a particle $i$ is located at the origin at time $t=0$ \cite{Hansen2013}. When particles act independently, this can be used to describe the probability density function describing motion of particles in the system (further explanation for this relationship is given in the Supplementary Materials).
 Assuming the imaged motion is isotropic and that the images are not spatially heterogeneous, an azimuthal average can be applied to Equation \eqref{eq: DDM_Decomp} for dimensional reduction and to average over random movement within the image stack. We express this as the contour integral \[\mathcal{D}(q,\Delta t) = \frac{1}{||\Gamma||}\oint_{\Gamma} \mathcal{D}(\bm{q},\Delta t) d\Gamma = \frac{1}{2 \pi }\int_{-\pi}^{\pi} \mathcal{D}(q,\theta,\Delta t) d\theta,\]
where $||\cdot||$ is the length of the contour $\Gamma$, and
\[\Gamma = q(\cos(\theta),\sin(\theta)), \quad \theta \in [0,2 \pi).\]
The azimuthal average of $\mathcal{D}(\bm{q},\Delta t)$ is a matrix, commonly expressed as
\begin{align}
	\mathcal{D}(q,\Delta t) = A(q) (1-f(q,\Delta t)), \label{eq: Minimise_funct}
\end{align}
where \[A\left(q\right) = \frac{2}{||\Gamma||} \oint_{\Gamma} \left\langle\left|\tilde{I}^2(\bm{q},t)\right|\right\rangle_t d\Gamma,\] is a scaling parameter depending on the appearance of particles \cite{Cerbino2008}. In practice, it is necessary to add an additional term $B(\bm{q})$ to Equation \eqref{eq: Minimise_funct}, which accounts for random noise from the camera \cite{Germain_2016, Drechsler_2017, Martinez_2012}. If an analytic form of $f(q,\Delta t)$ is known, we can evaluate $\mathcal{D}(q,\Delta t)$ numerically by iterating Equation \eqref{eq: DDM} over the image stack, and then employ numerical fitting approaches to Equation \eqref{eq: Minimise_funct} to determine parameters within $f(q,\Delta t)$. For example, in Brownian motion, it is known that $f(q,\Delta t)$ has the form

\[f(q,\Delta t) = e^{-q^2 D \Delta t},\]
where $D$ ($\diffunit$) is the diffusion coefficient \cite{Cerbino2008, Germain_2016, Berne2000}. If we fix $q$ in the DDM matrix, both $A(q)$ and $B(q)$ in Equation \eqref{eq: Minimise_funct} become constant, allowing us to first probe the time dependency of $f(q,\Delta t)$, and then fit again to probe the dependency on $q$. Often for diffusion problems \cite{Cerbino2008,Germain_2016,Drechsler_2017}, this two step approach first numerically fits some $q$ dependent term $\tau_q =(q^2 D)^{-1}$, and then determines $D$ by fitting
\[\log(\tau_q) = -2 \log(q) - \log(D).\]

\subsection*{DIC microscopy induces anisotropy in the DDM tensor}
\label{section: anisotropy_proof}
\begin{figure}[!htb]
	\centering
\begin{subfigure}[h]{0.49\linewidth}
	\centering
	\includegraphics[width =0.8\linewidth]{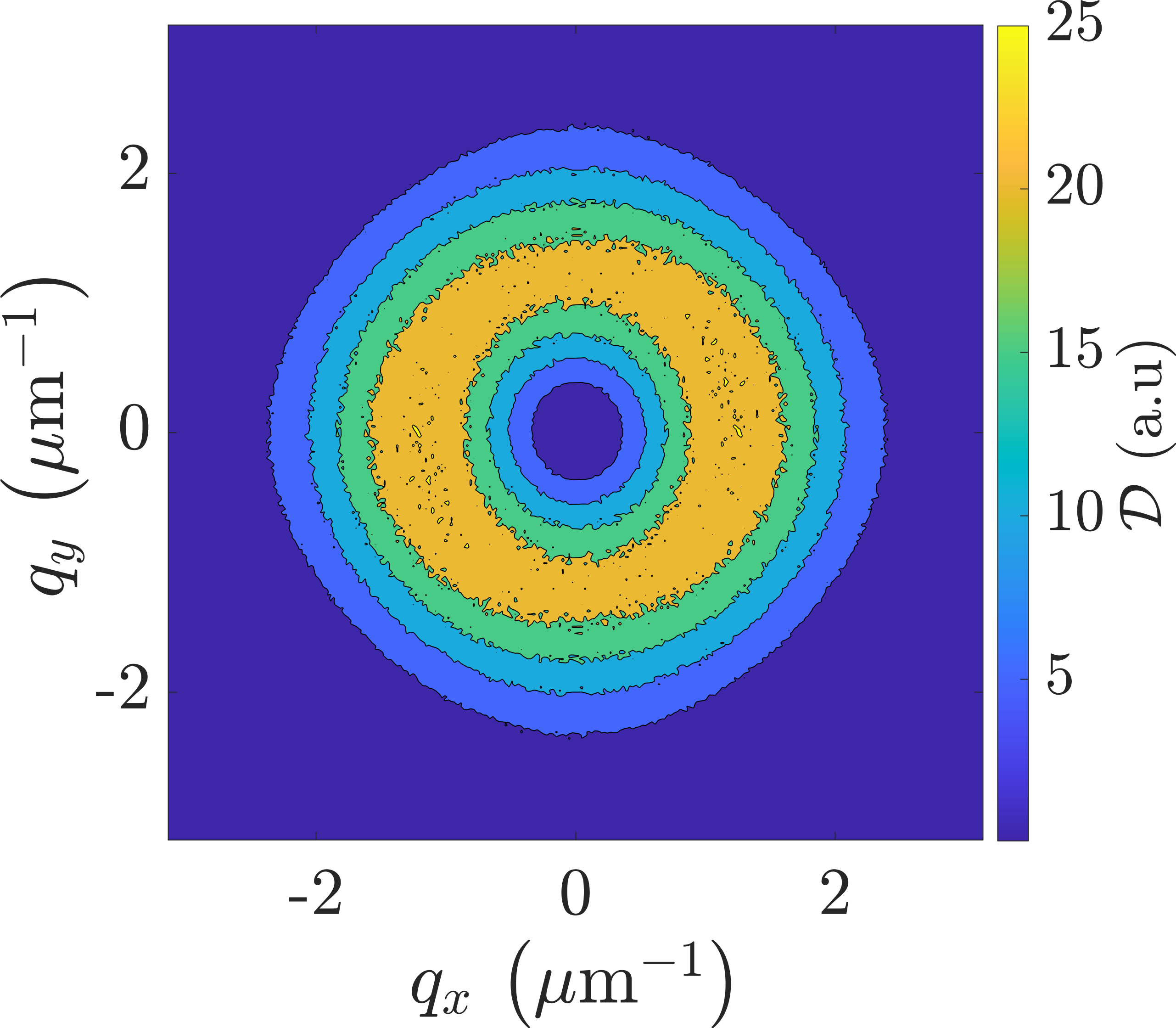}
	\caption{Bright field \cite{Germain_2016}.}
	\label{subfig: Brightfield}
\end{subfigure}
	~
\begin{subfigure}[h]{0.49\linewidth}
	\centering
	\includegraphics[width =0.8\linewidth]{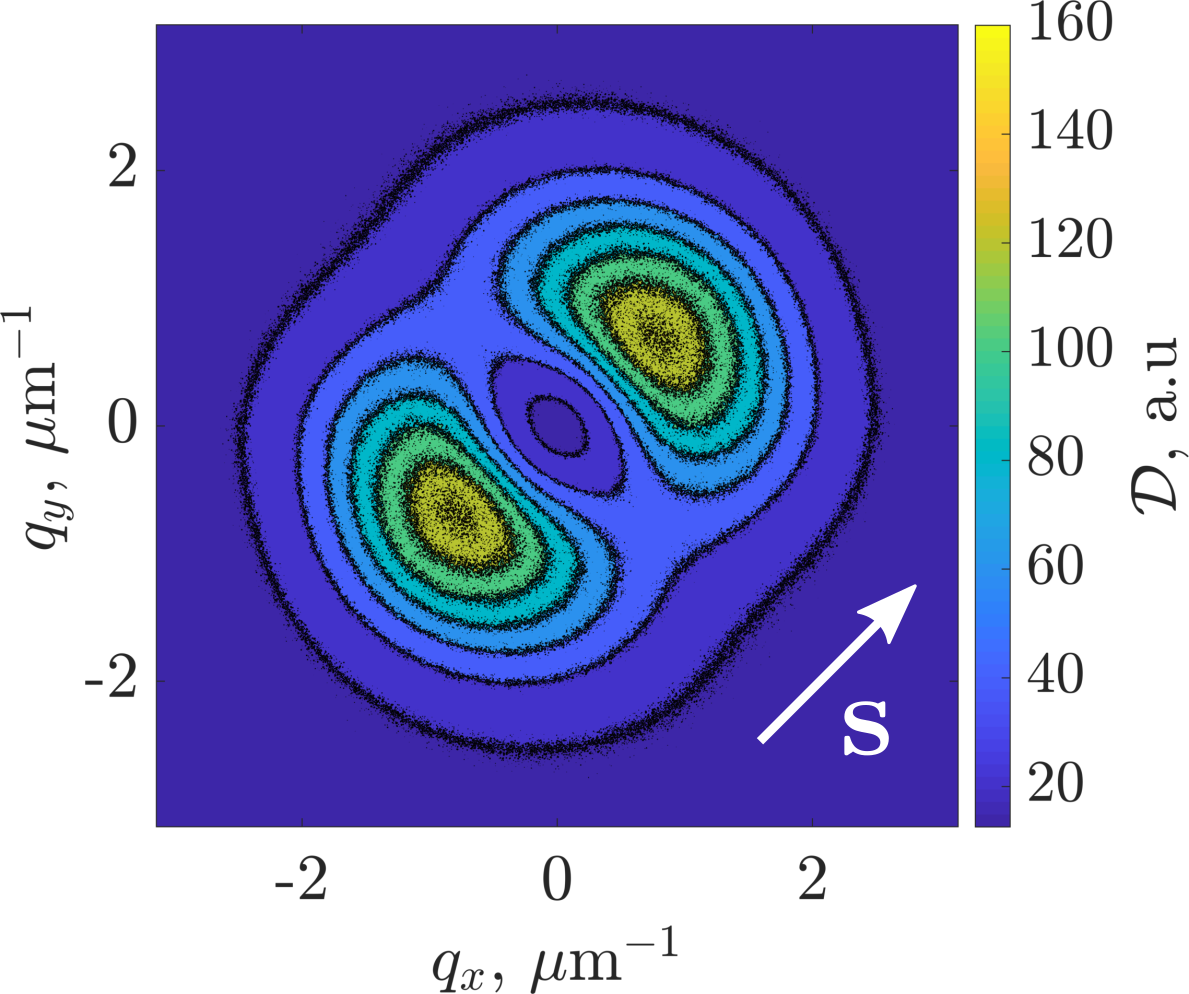}
	\caption{DIC.}
	\label{subfig: DIC}
\end{subfigure}
\caption{Isocontours of $\mathcal{D}(\bm{q},\Delta t)$ for \ref{subfig: Brightfield} bright field images of colloidal dynamics \cite{Germain_2016} and \ref{subfig: DIC} DIC images of colloidal dynamics. Both sets of images are displayed for the smallest available time separation: $\Delta t = 2.5$ ms for bright field and $0.12$s for DIC. The bright field images yield concentric circular isocontours, but in DIC images, there is a clear asymmetry aligned with the direction of the DIC microscope shear, denoted by the white arrow \bs.}
\label{fig: Isocontours}
\end{figure}

In Figure \ref{fig: Isocontours}, we plot $\mathcal{D}(\bm{q},\Delta t)$ at the smallest time lag available for two different movies of colloidal dispersions; \ref{subfig: Brightfield} comes from bright field images of a colloidal dispersion \cite{Germain_2016}, with concentric rings demonstrating that the underlying process is isotropic \cite{Germain_2016}. However, \ref{subfig: DIC}, derived from DIC images of a similar colloidal dispersion, shows an asymmetric `kidney bean' shape, where the highest peaks align along the orientation of the shear $\bs$ in the DIC image, and the lowest are orthogonal to it.  

Since we know the colloidal dispersion is subject only to isotropic Brownian motion \cite{Poon_2015}, the anisotropy in Figure \ref{fig: Isocontours}, right, is an artefact of the DIC imaging process. We now derive an expression to characterise and resolve this anisotropy.

\subsection*{DIC images as the difference of two phase images}
\label{section: I=Y-X}
A single DIC image can be expressed as the difference between overlapping identical offset images \cite{Shribak2017}, where each polarised beam in the pair generates a different image. For a DIC image stack $I(\bm{r},t)$, there exists image stacks $X(\bm{r},t)$ and $Y(\bm{r},t)$ such that 
\begin{align*}
	I(\bm{r}, t) &= Y(\bm{r},t) - X(\bm{r},t),\\
	Y(\bm{r},t) &= X(\bm{r} + \bs,t), 
\end{align*}
for a shear $\bs = (x_0, y_0)$. For some initial time, $t$, and some time displacement $\Delta t$, let $I_2 = I(\bm{r},t+\Delta t)$ and $I_1 =I(\bm{r},t)$, with the relation $I_j = Y_j - X_j,$ $ j \in \{1,2\}$. Hence, $\Delta I=\Delta Y - \Delta X$, and we construct the DDM matrix for the DIC image, denoted $\mathcal{D}_I$, as follows;
\begin{align}
	\mathcal{D}_I(\bm{q},\Delta t)	&= \left\langle \left|\Delta \tilde{Y}\right|^2 \right\rangle_t+ \left\langle \left|\Delta \tilde{X}\right|^2 \right\rangle_t- \left \langle \Delta \tilde{Y}\Delta \tilde{X}^*\right \rangle_t - \left \langle \Delta \tilde{X}\Delta \tilde{Y}^* \right \rangle_t. \label{eq: DDM_construction}
\end{align}

Comparing to Equation \eqref{eq: DDM}, $\mathcal{D}_Y \equiv \left\langle \left|\Delta \tilde{Y}\right|^2 \right\rangle_t$ and $\mathcal{D}_X \equiv \left\langle \left|\Delta \tilde{X}\right|^2 \right\rangle_t$. In DIC microcopy, shear distances may be as small as 0.2-0.3 $\mu$m \cite{Regan2019,Shribak2017,Murphy_2012}, which means that this displacement is small relative to the size of the image. Hence, it is unlikely that the sampled dynamics between the image sets will differ significantly, and we conclude that $\mathcal{D}_Y \approx \mathcal{D}_X$.

By the translation property of the Fourier transform,
\begin{align}
	\tilde{Y_j}(\bm{q}) = \tilde{X_j}(\bm{q}) e^{-i \left(\bm{q} \cdot \bs \right)},\label{eq: single_fourier_shift}
\end{align}

and rearranging the definition of the ISF in \eqref{def: ISF}, we may rewrite Equation \eqref{eq: DDM_construction} as
\begin{align*}
	\langle|\Delta \tilde{I}|^2\rangle_t &=  2 \mathcal{D}_X(\bm{q},\Delta t) -4|\tilde{X}|^2  \cos\left( \bm{q}\cdot \bs\right)\left(1 - f(\bm{q},\Delta t)\right). 
\end{align*}
Assuming a sufficiently high signal to noise ratio, such that noise is negligible, Equation \eqref{eq: Minimise_funct} defines the DDM tensor on a DIC image stack as
\begin{align}
	\mathcal{D}_I(\bm{q},\Delta t) = 4|\tilde{X}|^2(1-f(\bm{q},\Delta t))(1-\cos(\bm{q} \cdot \bs)). \label{eq: DIC_DDM_Beforeaverage}
\end{align}

As previously, the next step in the DDM analysis is to apply an azimuthal average. We first simplify our analysis by assuming the underlying motion, and therefore $f(\bm{q},\Delta t)$, is isotropic, for example in the case of swimming bacteria \cite{Wilson_2011} or colloidal dispersions \cite{Cerbino2008}.

\section{Fitting invariance for isotropic ISF}
\label{section: isotropic}
Assuming that $f(\bm{q},\Delta t)$ is isotropic, and noting that any circular contour along $\mathcal{D}(\bm{q},\Delta t)$ centered on $\bm{q}=(0,0)$ is periodic with period $\pi$, 
the azimuthal average of \eqref{eq: DIC_DDM_Beforeaverage} can be written as
\begin{align*}
	\mathcal{D}_I(q,\Delta t) = \oint_{\Gamma}	\mathcal{D}_I(\bm{q},\Delta t)d\Gamma	&=	(1-f(\bm{q},\Delta t))\frac{1}{|\Gamma|}\oint_{\Gamma} 4 |\tilde{X}|^2 (1-\cos(\bm{q}\cdot \bs)).
\end{align*}
Defining some new scaling term depending only on $q$ by
\[A_{I}(q) = \frac{1}{|\Gamma|}\oint_{\Gamma} 4 |\tilde{X}|^2 (1-\cos(\bm{q}\cdot \bs)),\]
we can express $\mathcal{D}_I(q,\Delta t)$ in the form
\begin{align}
	\mathcal{D}_I(q,\Delta t) =  A_I(q)(1-f(q,\Delta t)). \label{eq: Rescaled DDM}
\end{align}
Hence, the functional form of the ISF remains unchanged, despite the shadow in DIC images, and hence running parameter fitting on either $\mathcal{D}_X$ or $\mathcal{D}_I$ will yield equivalent motility parameters. A more specific form of $A_I(q)$ can be found when $ |\tilde{X}|^2$ does not depend on the orientation of $\bm{q}$, a realistic condition which holds for collections of well mixed particles that share no common orientation. Under this condition, by the definition of the Bessel function of order zero \cite{Abramowitz1964},
\[J_0(x) = \frac{1}{\pi}\int_0^{\pi} \cos\left(x \cos(\theta)\right) d\theta,\]
the scaling term reduces to 
\begin{align}
	A_I(q) = 4 |\tilde{X}|^2(1-J_0(q s)) = 2 A_X(q) \left(1-J_0(q s)\right),\label{eq: Amap}
\end{align}

where $s=|\bs| = \sqrt{x_{0}^2 + y_0^{2}}$. Thus, we express the radially averaged DDM matrix for the DIC image as
\begin{align}
	\mathcal{D}_I(q,\Delta t) =2(1-J_0(q s)) \mathcal{D}_X(q,\Delta t). \label{eq: Map_x_to_I}
\end{align}
Hence, $\mathcal{D}_I(q,\Delta t)$ is a rescaled version of $\mathcal{D}_X(q,\Delta t)$, implying that DDM is robust even under the shadowing effect leading to the deformation of the DDM matrix in DIC images. We now validate this expression, through both simulated datasets and real DIC images.
\subsection*{Validation in simulations and colloid data}
\begin{figure}[H]
	\centering
	\includegraphics[scale=.4]{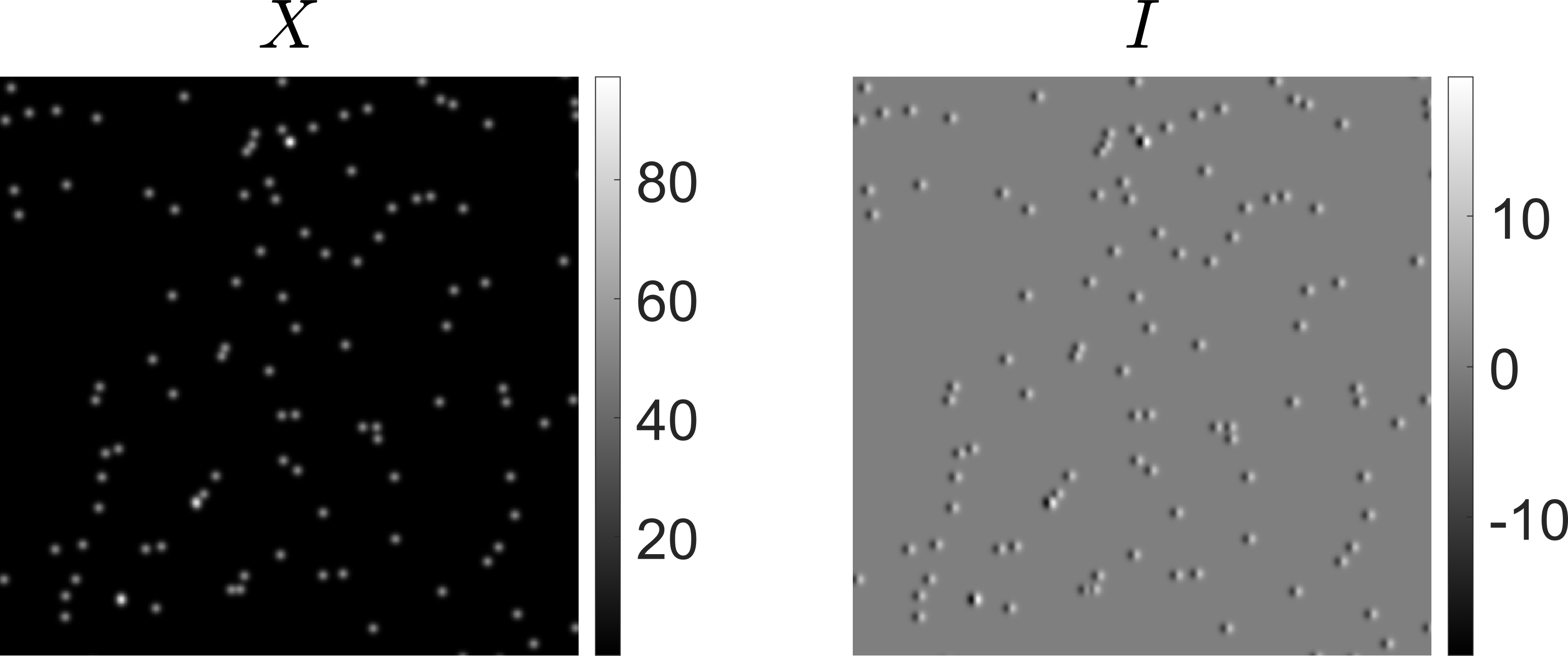}
	\caption{The first frame of the simulated phase image, $X$, and the corresponding DIC image, $I$, which is generated by $X(\bm{r}+\bm{s},t)$-$X(\bm{r},t)$.}
	\label{fig: firstframe}
\end{figure}

We generate simulations of 150 particles moving inside a 490$\times$490 window, with diffusion coefficient $D=0.5\mu \text{m}^2/\text{s}$, where 1 pixel represents 1 $\mu$m, for a duration of 4000 frames at a frequency of $8$ frames per second. From these simulations, we generate the phase image stack $X(\bm{r},t)$ and the corresponding DIC image stack $I(\bm{r},t)$ using a DIC shift of 1 pixel to the right, i.e $\bm{s}=(1,0)$. The first frame of each image stack is visualised in Figure \ref{fig: firstframe} From each image stack, we generate the corresponding DDM matrices $D_X$ and $D_I$, respectively. More details about the simulations are available in Supplementary Material, and the code is available at \url{https://github.com/OstlerT/DIC_DDM_Ostler.git}.

\begin{figure}[ht!]
	\centering
	\includegraphics[width=0.8 \linewidth]{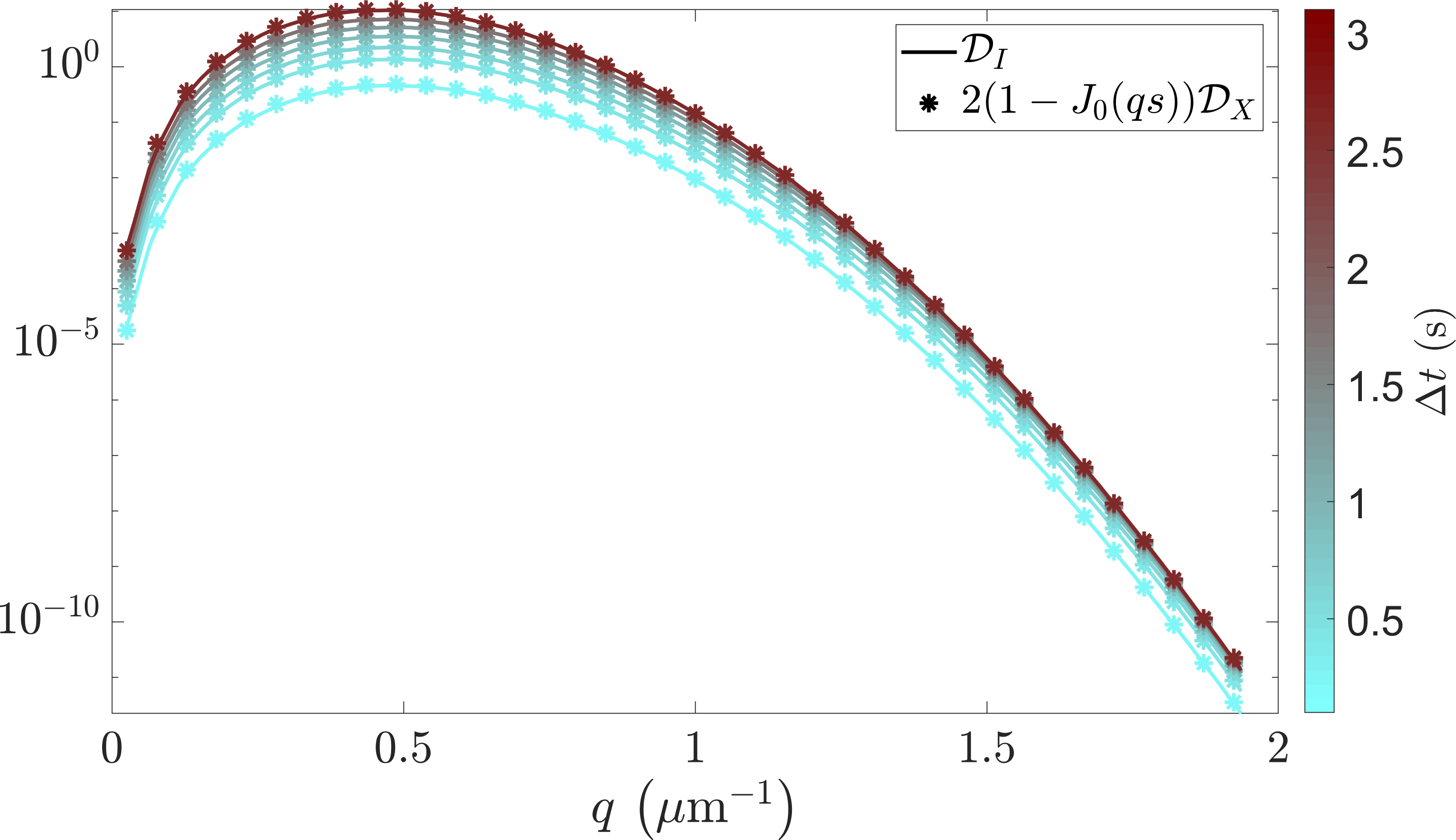}
	\caption{Snapshots of $\mathcal{D}_I(q,\Delta t)$ and the rescaled form $2(1-J_0(q s))\mathcal{D}_X(q,\Delta t)$, plotted at fixed times over all spatial frequencies. There is excellent agreement for all frequencies $q < 2 \mu\text{m}^{-1}$ hence the relationship \eqref{eq: Map_x_to_I} is valid over this interval. }
	\label{fig: DDMrelation}
\end{figure}

In Figure \ref{fig: DDMrelation}, we demonstrate that the decomposition in Equation \eqref{eq: Map_x_to_I} holds for all frequencies $q < 2 \mu\text{m}^{-1}$. For different sets of simulation parameters, the agreement can deteriorate at very high and very low frequencies, but it is standard practice in DDM analysis to restrict fitting to some interval on $q$ away from the lowest available frequencies, where statistical sampling of the radial average is poor, and away from large frequencies where the signal to noise ratio is poor \cite{Germain_2016}. As such, as long as the relationship holds over the optimal fitting region, we can conclude that the analysis of $\mathcal{D}_X$ and $\mathcal{D}_Y$ will yield equivalent results.

\begin{figure}[!htb]
	\centering
\begin{subfigure}[h]{0.8\linewidth}
	\centering
	\includegraphics[width =0.7\linewidth]{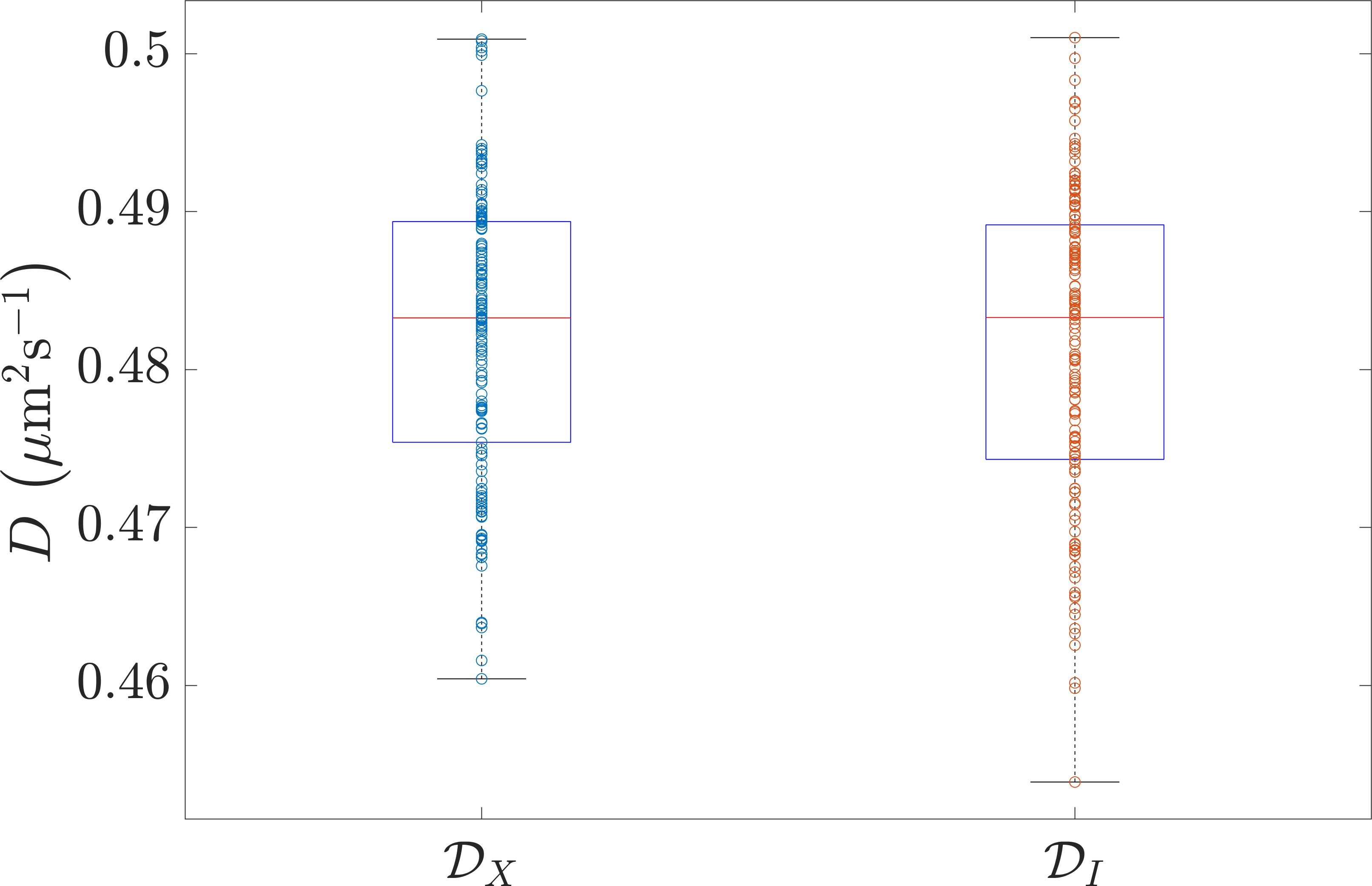}
	\caption{Distributions of the diffusion coefficients fitted from $\mathcal{D}_X$. and $\mathcal{D}_I$. }
	\label{subfig: Boxplots}
\end{subfigure}
	~
\begin{subfigure}[h]{0.8\linewidth}
	\centering
	\includegraphics[width =0.7\linewidth]{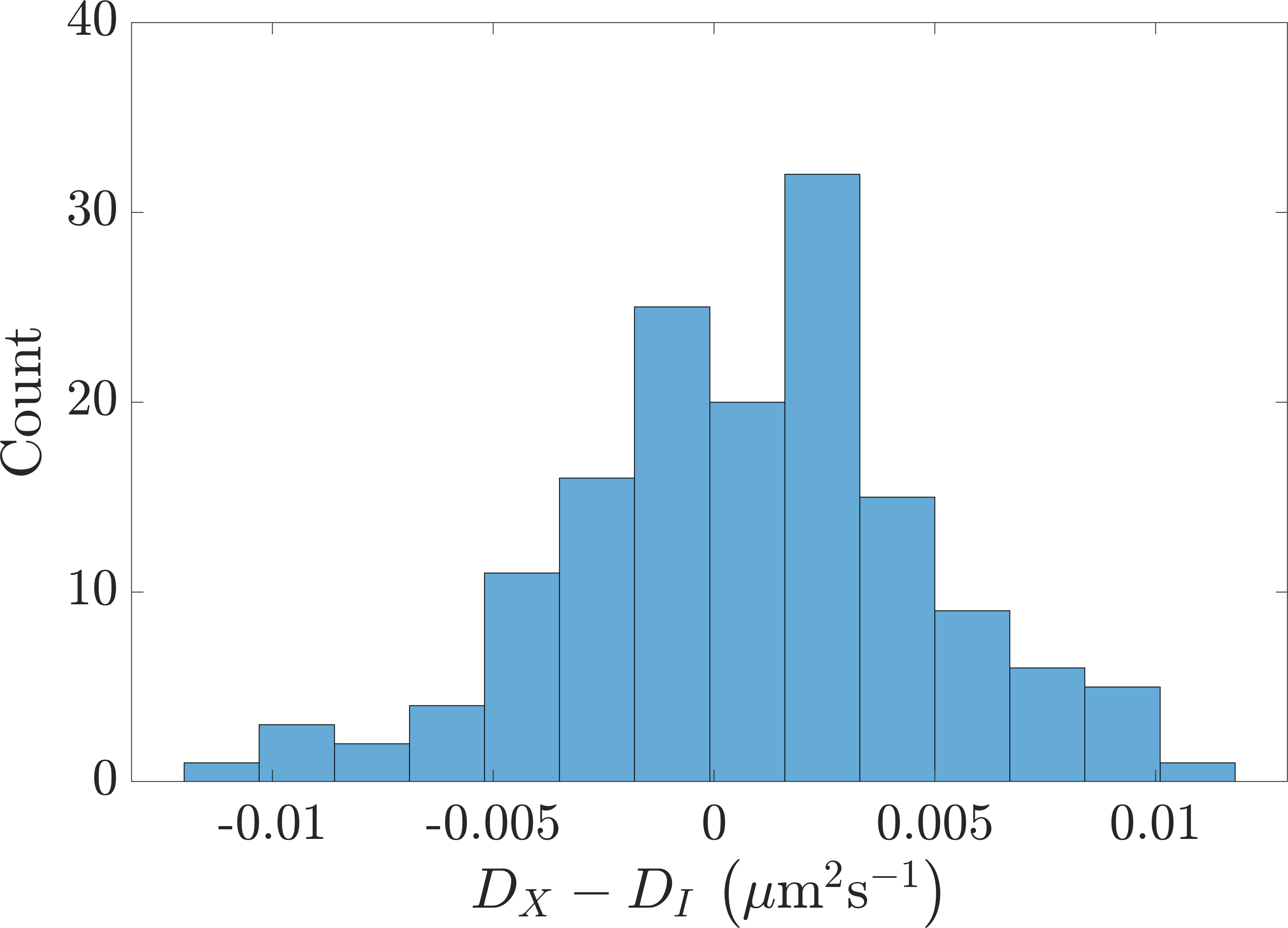}
	\caption{Distribution of the difference $D_X-D_I$ for each simulation, with mean $6.9 \times 10^{-4}$. }
	\label{subfig: histogram}
\end{subfigure}
\caption{For 150 simulations of Brownian motion, we generate phase stacks $X(\bm{r},t)$ and corresponding DIC images $I(\bm{r},t)$. DDM analysis on each stack yields a diffusion coefficient, $D_X$ and $D_I$, respectively. \ref{subfig: Boxplots} Although the fitted diffusion coefficients have some spread around the true value $D=0.5 \diffunit$, \ref{subfig: histogram} the difference $D_X-D_I$ is small, so the DIC shear has minimal effect on the fitting accuracy. }
\label{fig: sdescribe}
\end{figure}

We now show that the diffusion coefficients generated by fitting $\mathcal{D}_X$ and $\mathcal{D}_I$ are equivalent. We generate 150 movies of diffusing particles, choosing the diffusion coefficient to be $D=0.5\, \mu \text{m}^2 /\text{s}$ to remain consistent with literature simulation approaches \cite{Bayles_2016}, and perform fitting over the frequency interval $0.3\mum<q<0.8\, \mum$. The average diffusion coefficients generated by the simulations of $X$ and $I$ are $\overline{D}_X =0.481 \pm 0.009 \diffunit$, and $\overline{D}_I = 0.482 \pm 0.010\diffunit$, respectively, with the distribution of the fitted diffusion coefficients plotted in Figure \ref{subfig: Boxplots}. Although a wide spread of fitted diffusion coefficients arises in the simulations, plotting the values $D_X-D_I$ as a histogram in Figure \ref{subfig: histogram} shows the fitting difference is small between $\mathcal{D}_X$ and $\mathcal{D}_I$. Hence, DIC shadowing has a minimal effect on the fitted diffusion coefficient.

\begin{figure}[ht!]
	\centering
	\includegraphics[width=.9\linewidth]{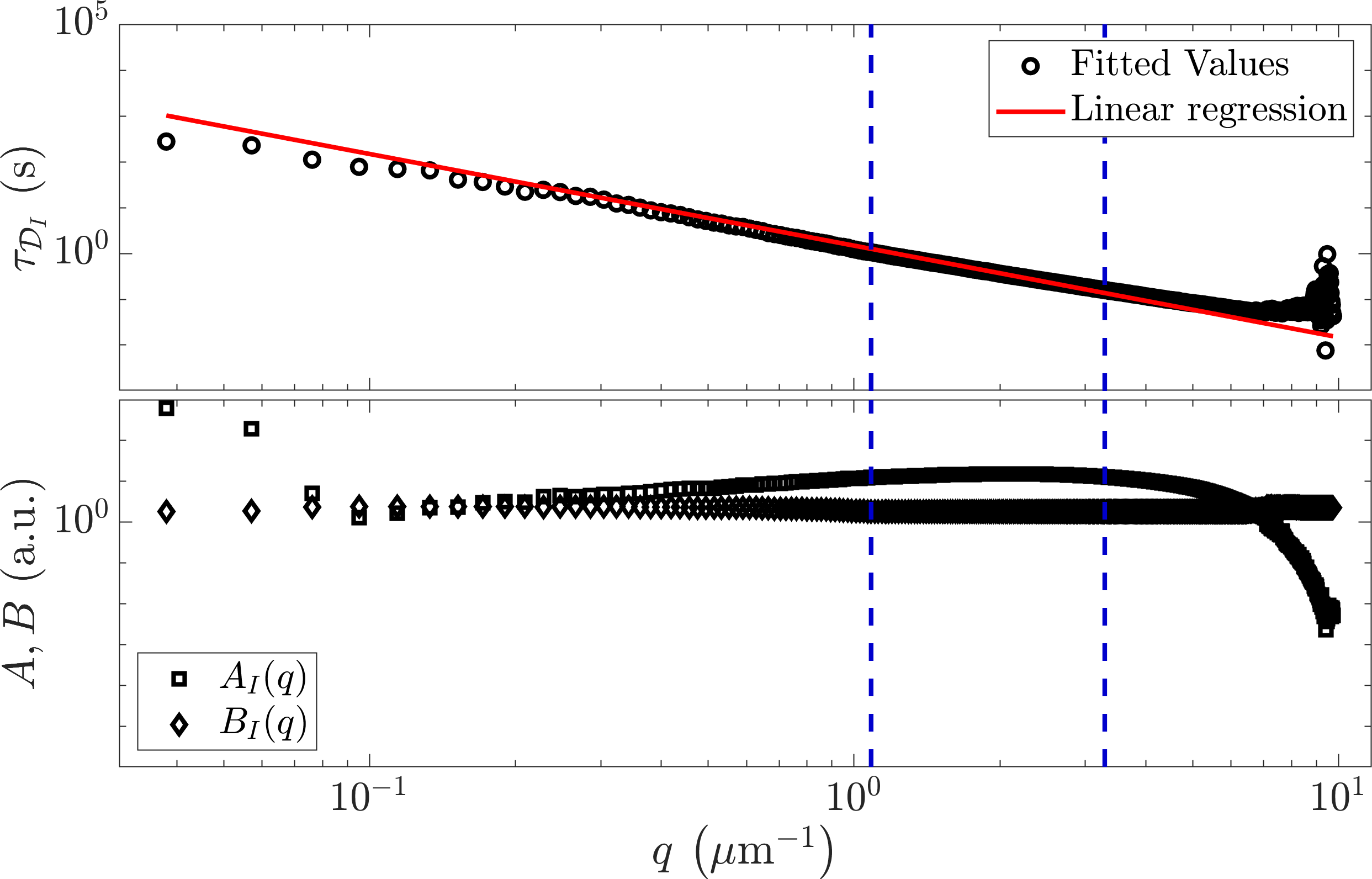}
	\caption{DDM analysis of the diffusion coefficients for fitted colloidal DIC data, yielding a fitted diffusion coefficient of $0.676 \diffunit$. The fitting region $1.1\, \mum<q<3.3\, \mu \text{m}^{-1}$ , marked between the vertical blue dashed lines, is chosen where the signal ($A_I(q)$) to noise ($B_I(q)$) ratio is sufficiently high. }
	\label{fig: Realdatafits}
\end{figure}

Additionally, we show that DDM is accurate in a single image stack formed of 7680 DIC images of a colloidal dispersion, as described in Section \ref{sec:Samples}. The images are $1344\times 1024$ pixels representing a $433.44\times 330.24$ $\mu$m real region (pixel length $322.5$ nm), from which we take a square subset with a side length of $1024$ pixels. The shearing distance is measured to be $238 \pm 10$ nm \cite{Regan2019}, the pixel length is $322.5 \,$nm, with a shear angle of $\phi = \pi/4$ and a phase offset of 90$^{\circ}$. From the corresponding DDM matrix, $\mathcal{D}_I$, we fit a diffusion coefficient of $0.676 \, \diffunit$.
Comparing this to the Stokes-Einstein relation \cite{Cerbino2008,Germain_2016},
\[D=\frac{K_{B}T}{6 \pi \eta r},\]
the expected diffusion coefficient is $0.667$ $\diffunit$, where $K_B$ is the Boltzmann constant, $T=291.25$K is absolute temperature, $\eta=3.20$ mPa $(\text{s}^{-1})$ is the dynamic viscosity of a $30\%$ glycerol solution, and $r$ the particle radius. Hence, $D$ is comparable to the fitted value. We visualise $\tau_{\mathcal{D}_I}$, $A(q)$, $B(q)$ and the fitting regime in Figure \ref{fig: Realdatafits}.

Although we have only shown that our results hold for Brownian motion, the relationship \eqref{eq: Amap} holds for all forms of isotropic motion. However, we cannot make the same claims for anisotropic motion, since the azimuthal average of Equation \eqref{eq: DIC_DDM_Beforeaverage} contains the term
\[\oint_{\Gamma}f(\bm{q},\Delta t) \cos(\bm{q} \cdot \bm{s}), \]
which we cannot evaluate without knowing the exact form of $f(\bm{q},\Delta t)$. For now, we consider directed advection-diffusion, a specific form of anisotropy present in the motivating example of classifying the rate of cytoplasmic movement in oocytes \cite{Drechsler_2017}.
\section{Exploring anisotropy; Fitting invariance for an advection-diffusion process}
\label{Section: AdvectionDIC}
In order to distribute vesicles carrying important cell content, directed transport occurs in the cytoplasm of the oocyte, which induces a large-scale net flow known as cytoplasmic streaming \cite{Drechsler_2017}. This process results in particles having a constant, directed velocity, superimposed on the diffusive random processes. The probability density function $u(\bm{r},\Delta t)$ describing the distribution of displacements of particles as they diffuse and move with velocity $\bm{v}$ satisfies the advection-diffusion equation,
\[\D{u(\bm{r},\Delta t)}{(\Delta t)} = D \nabla^2 u(\bm{r},\Delta t) - \bm{v} \cdot \nabla u(\bm{r},\Delta t),\]
where $D$ is the diffusion coefficient as in Section \ref{section: DDM}, and $\bm{v}$ $\left(\mu \text{m}^{-1}\right)$ is assumed to be constant. The corresponding ISF is \cite{Giavazzi_2014}
\[f(\bm{q},\Delta t) = e^{-q^2 D \Delta t} \cos(\bm{q} \cdot \bm{v} \Delta t).\]
The formulation of the DDM matrix for DIC images, advection-diffusion processes, follows the same argument as in Section \ref{section: I=Y-X}, up to Equation \eqref{eq: DIC_DDM_Beforeaverage}. Substituting the new ISF into the definition of $\mathcal{D}_I(\bm{q},\Delta t)$ yields
\begin{align*}
	\mathcal{D}_I (\bm{q},\Delta t) &= 4|\tilde{X}|^2 (1-e^{-q^2 D \Delta t} \cos(\bm{q} \cdot \bm{v} \Delta t) - \cos(\bm{q}\cdot \bs) + e^{-q^2 D \Delta t} \cos(\bm{q} \cdot \bm{v}\Delta t)\cos(\bm{q}\cdot\bs))).
	%&= 4|\tilde{X}|^2 \left( 1 - \cos(\bm{q} \cdot \bs)\right) \left( 1-e^{-q^2 D \Delta t}\cos \left( \bm{q} \cdot \bm{v} \Delta t\right)\right). \nonumber
\end{align*}
We convert to polar coordinates, using
\begin{align*}
	\bm{v} &= v\left(\cos \phi_1 ,\sin\phi_2\right),\\
	\bs &= s\left(\cos\phi_2,\sin\phi_2\right),\\
	\bm{q} &= q\left(\cos\theta,\sin\theta\right),\\
	\text{and } \frac{1}{||\Gamma||}\oint_{\Gamma} \mathcal{D}(\bm{q},\Delta t) d \Gamma &= \frac{1}{2 \pi q} \int_{-\pi}^{\pi}\mathcal{D}(q,\theta,\Delta t) q d\theta = \frac{1}{2 \pi } \int_{-\pi}^{\pi} \mathcal{D}(q,\theta,\Delta t) d\theta.
\end{align*}

\begin{figure}[ht!]
	\centering
	\includegraphics[width=0.8 \linewidth]{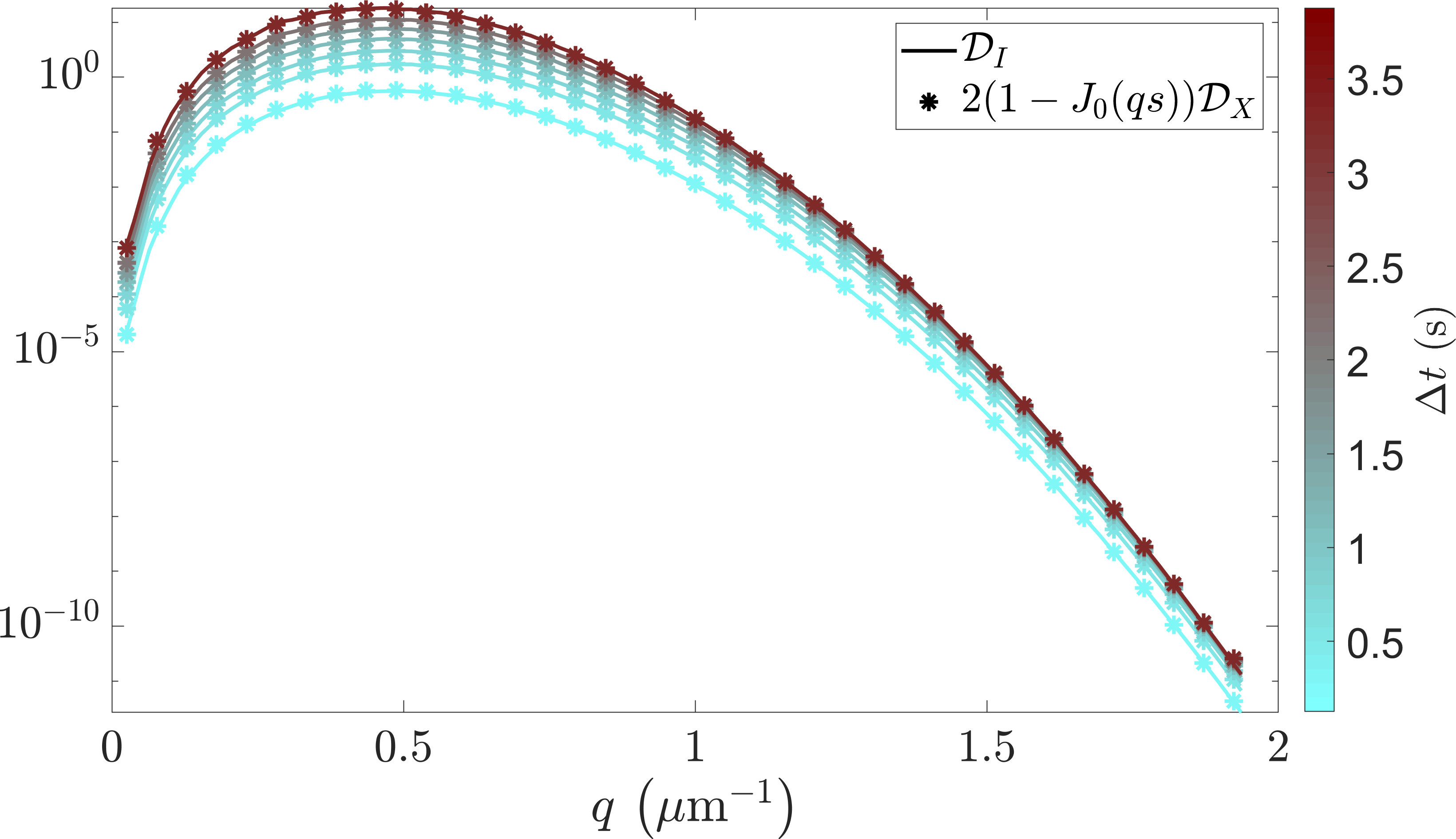}
	\caption{Snapshots of $\mathcal{D}_I(q,\Delta t)$ and the rescaled form $2(1-J_0(q s))\mathcal{D}_X(q,\Delta t)$, plotted at fixed times over all spatial frequencies, for an anisotropic advection-diffusion process. Where the shear is small relative to the displacement from advection, Relationship \eqref{eq: Map_x_to_I} holds.}
	\label{fig: DDMrelationvelocity}
\end{figure}

Since the contour on which we take the azimuthal average of $\mathcal{D}(q,\theta,\Delta t)$ is $\pi$-periodic, we restrict $\phi_1,$ $ \phi_2 \in [-\pi/2,\pi/2)$ and take the azimuthal average as
\begin{align}
	\mathcal{D}_I(q,\Delta t) =\frac{4|\tilde{X}|^2  }{ \pi}\int_{-\frac{\pi}{2}}^{\frac{\pi}{2}} &\left( 1-e^{-\kappa}\cos \left(\lambda\cos \left(\theta - \phi_1 \right) \right) - \cos \left(\xi \cos \left(\theta - \phi_2 \right) \right) \right. \nonumber \\
	&\left.+e^{-\kappa}\cos \left(\lambda\cos \left(\theta - \phi_1 \right)\right)\cos \left( \xi \cos \left(\theta-\phi_2 \right) \right)\right)  d \theta, \label{Eq: RadAvgDDMI}
\end{align}
where we have introduced the non-dimensional parameters
\begin{align*}
	\lambda= q v \Delta t, \quad \xi= q s, \quad \kappa = q^2 D \Delta t.
\end{align*}
We may rewrite the final term of the integral in Equation \eqref{Eq: RadAvgDDMI} as
\begin{align*}
	&\frac{e^{-\kappa}}{\pi}\int_{-\frac{\pi}{2}}^{\frac{\pi}{2}} \cos \left(\lambda\cos \left(\theta - \phi_1 \right)\right)\cos \left( \xi \cos \left(\theta-\phi_2 \right) \right)  d \theta \nonumber\\
	&=e^{-\kappa}\frac{1}{2\pi} \int_{-\frac{\pi}{2}}^{\frac{\pi}{2}} \cos(R_{+} \cos(\theta - \psi_{+})) d\theta + e^{-\kappa}\frac{1}{2\pi} \int_{-\frac{\pi}{2}}^{\frac{\pi}{2}} \cos(R_{-} \cos(\theta - \psi_{-})) d\theta, \nonumber 
\end{align*}
where 
\begin{align*}
	R_{\pm} \cos(\theta - \psi_{\pm}) &= R_{\pm} \cos(\theta) \cos(\psi_{\pm}) + R_{\pm} \sin(\theta)\sin(\psi_{\pm}),\\
	\text{and } R_{\pm}&=\sqrt{\lambda^2 + \xi^2 \pm 2 \lambda \xi \cos\left(\phi_1-\phi_2\right)}, \quad \psi_{\pm} =  \arctan\left(\frac{\lambda \sin(\phi_1) \pm \xi \sin(\phi_2)}{\lambda \cos(\phi_1) \pm \xi \cos(\phi_2)}\right).
\end{align*}
Noting that $\psi_{\pm}$ does not depend on $\theta$, we may express the integral in Equation \eqref{Eq: RadAvgDDMI} using Bessel functions of order 0,
\begin{align}
	\mathcal{D}_I &= 4|\tilde{X}|^2 \left( 1 - e^{-\kappa}J_0(\lambda) - J_0(\xi) + \frac{e^{-\kappa}}{2}\left(J_0 \left(R_{+}\right) + J_0\left(R_{-}\right)\right)\right), \label{eq: DIC_Advection_form}
\end{align}
which gives us a new ISF,
\begin{align}
	f(q,\Delta t) &= J_0(\xi) + e^{-\kappa}J_0(\lambda)  \nonumber\\
	&-\frac{e^{-\kappa}}{2}\left(J_0\left(\sqrt{\lambda^2 + \xi^2 + 2 \lambda \xi \cos(\phi_1 - \phi_2)}\right) + J_0\left(\sqrt{\lambda^2 + \xi^2 - 2 \lambda \xi \cos(\phi_1 - \phi_2)}\right)\right). \label{eq: New_ISF}
\end{align}
Unlike the case of isotropic motion, the ISF now depends on the DIC shear parameters $\xi$ and $\phi_2$, and we cannot immediately conclude that the DDM analysis on DIC images will yield accurate fitting parameters for an advection-diffusion process. However, we can use the fact that the DIC shear is small relative to the advective displacements to obtain an approximation to Equation \eqref{eq: New_ISF}, and extend our isotropic conclusion to advection-diffusion processes. Consider the term 
\[h(\lambda, \xi, \phi_1,\phi_2) = \frac{1}{2}\left(J_0\left(\sqrt{\lambda^2 + \xi^2 + 2 \lambda \xi \cos(\phi_1-\phi_2)}\right) + J_0\left(\sqrt{\lambda^2 + \xi^2 - 2 \lambda \xi \cos(\phi_1-\phi_2)}\right)\right), \]
and define
\begin{align*}
g(\lambda,\xi) = J_0(\lambda)J_0(\xi),
\end{align*}
which we will evaluate as an approximation to $h(\lambda, \xi, \phi_1,\phi_2)$. A Taylor series of both $h$ and $g$ about $\xi = 0$ becomes
\begin{align*}
	h(\lambda,\xi,\phi_1,\phi_2) &= J_0(\lambda) + \frac{1}{4} \left(J_2 \left(\lambda\right)\left( \cos(2 (\phi_1-\phi_2))\right) - J_0(\lambda)\right)\xi^2 + \mathcal{O}(\xi^4),\\
	g(\lambda,\xi) &= J_0(\lambda) \left( 1-\frac{1}{4} \xi^2 \right)+\mathcal{O}(\xi^4).
\end{align*}
Thus, the difference between $h(\lambda,\xi,\phi_1,\phi_2)$ and $g(\lambda,\xi)$ for small $\xi$ is approximately
\begin{align*}
	h(\lambda,\xi,\phi_1,\phi_2)-g(\lambda,\xi) = \frac{\xi^2}{4}J_2(\lambda)\left(\cos\left(2 (\phi_1-\phi_2)\right)\right) + \mathcal{O}(\xi^4).
\end{align*}
 Therefore the functions $h$ and $g$ are approximately equal, since $J_2(\lambda)<1/2$ \cite{Mecholsky_2021} and, thus, the error term is bounded above by $\xi^2/8$. So, for sufficiently small $\xi$ we may rewrite the DDM matrix as
\begin{align*}
	\mathcal{D}_I &= 4|\tilde{X}|^2 \left(1-e^{-\kappa}J_0(\lambda) - J_0(\xi) + e^{-\kappa}J_0(\lambda)J_0(\xi)\right), \nonumber \\
	&= 4|\tilde{X}|^2 \left(1-e^{-\kappa} J_0(\lambda)\right)\left(1-J_0(\xi)\right), 
\end{align*}
which suggests that essentially the same DDM analysis can be applied to DIC images for anisotropic images without consideration of the DIC shear, since the effect of the DIC shear is accounted for entirely by a reparametrisation of the scaling term. We demonstrate this numerically, in Figure \ref{fig: DDMrelationvelocity}, by generating one simulation as in Section \ref{section: isotropic}, with the addition of a directed velocity of $v=0.5 \mu \text{m} \text{s}^{-1}$ orthogonal to the DIC shear direction, and plotting the remapped DDM matrix $\mathcal{D}_X(q,\Delta t)$ against $\mathcal{D}_I(q,\Delta t)$. Our choice of $v$ means that the displacement from velocity is equal to the DIC shear when $\Delta t = 2$s, which occurs after 16 frames out of the potential 4000, so the small DIC shear approximation is satisfied over almost all times.

We now summarise the conclusions of this work, and outline potential future developments.
\section{Discussion}
\label{Section: Further_Work}
We have shown that applying Differential Dynamic Microscopy (DDM) to Differential Interference Contrast (DIC) images results in deformation of the DDM matrix as a result of the characteristic shadowing of contrast microscopy. We also show that, for isotropic applications, this deformation can be described through multiplication by a time-independent scaling factor, such that the fitted output statistics from DDM remain invariant under the deformation. This result agrees with the results observed in \cite{Drechsler_2017}, in which DDM applied to isotropic regions of the Drosophila oocyte yielded motility parameters consistent with Particle Image Velocimetry (PIV). 

The power of this result lies in its simplicity for future applications of DDM; when studying isotropic motion in DIC images, no consideration is required to account for the DIC shadow, which from first inspection of the DDM tensor, might have appeared to have significant impact on the fitting of motility parameters. We highlight that one immediate application of this result is in the IVF clinic, since we validated that DDM would be robust in the phase-contrast images of human oocytes available to clinics, enabling the exploration of DDM as a tool non-invasive assessment of oocyte health in the clinic. However, our conclusions hold for any application of DDM.

We have additionally extended our analysis to advection-diffusion processes, under the realistic assumption that the rate of advection is much larger than the size of the DIC shear. This is a reasonable physical assumption, since the DIC shear is small by necessity. However, this does pose the question of what might happen when the shear and displacement from advection are of a similar order of magnitude. Exploratory simulations suggest that effect of the DIC shadow is negligible for any combination of shear displacement and velocity, such that DDM is always robust, but a more thorough analysis is required.

Since the only anisotropic form of motion we have considered is advection-diffusion, it would be tenuous to make more general comments about the interaction between the DIC shear and anisotropic motion. If there exists some form of motion for which the interaction between DDM and DIC is significant, further adjustment would be imperative during the fitting phase to guarantee accuracy. Unless we can prove that no form of anisotropic motion will significantly interact with DIC, potentially under some assumptions on the microscope setup and the underlying motion observed, caution should be exercised when applying DDM to anisotropic motion in DIC microscopy.

Although our results suggest that, in the case of DIC, no adjustment should be made for the application of DDM, there might be other forms of microscopy which have significant interactions with the output statistics of DDM. There is evidence of DDM being successfully applied in a variety of microscopy settings, but more in-depth mathematical investigation of the interaction between the output statistics of DDM, and the nuance of the imaging method, is more scarce. In Ref. \cite{Bayles_2016} it was shown that the inhomogeneous illumination inherent in dark field imaging will not significantly impact output statistics from DDM, under the condition that displacement is small relative to the gradient of the illumination, but an avenue for future research is to check the validity of the output statistics over the wide range of microscopy settings in different microscopy fields.
\section{Acknowledgements}
T.O acknowledges funding from Knowledge Economy Skills Scholarships (KESS2), a pan-Wales higher level skills initiative led by
Bangor University on behalf of the HE sector in Wales. It is part funded by the Welsh Government’s
European Social Fund (ESF) convergence programme for West Wales and the Valleys.
E.L recieved funding from the Engineering and Physical Sciences Research Council (EPSRC) ``Physics of Life'' Doctoral Training Partnership [project code EP/T517951/1].
Y.W was funded by the Wellcome trust.

	\bibliographystyle{vancouver}
	%\bibliography{biblio}

	\bibliography{biblio,WL}

\begin{thebibliography}{10}

\bibitem{Jepson2019}
Jepson A, Arlt J, Statham J, Spilman M, Burton K, Wood T, et~al.
\newblock {High-throughput characterisation of bull semen motility using
  differential dynamic microscopy}.
\newblock PLOS One. 2019;14(4):e0202720.

\bibitem{Ajduk_2011}
Ajduk A, Ilozue T, Windsor S, Yu Y, Seres KB, Bomphrey RJ, et~al.
\newblock {Rhythmic actomyosin-driven contractions induced by sperm entry
  predict mammalian embryo viability}.
\newblock Nat Commun. 2011;2(1).

\bibitem{Cerbino2008}
Cerbino R, Trappe V.
\newblock {Differential dynamic microscopy: probing wave vector dependent
  dynamics with a microscope}.
\newblock Phys Rev Lett. 2008;100(18):188102.

\bibitem{Wilson_2011}
Wilson LG, Martinez VA, Schwarz-Linek J, Tailleur J, Bryant G, Pusey PN, et~al.
\newblock {Differential Dynamic Microscopy of Bacterial Motility}.
\newblock Phys Rev Lett. 2011;106(1):018101.

\bibitem{Drechsler_2017}
Drechsler M, Giavazzi F, Cerbino R, Palacios IM.
\newblock {Active diffusion and advection in Drosophila oocytes result from the
  interplay of actin and microtubules}.
\newblock Nature communications. 2017;8(1):1-11.

\bibitem{He2012}
He K, Spannuth M, Conrad JC, Krishnamoorti R.
\newblock {Diffusive dynamics of nanoparticles in aqueous dispersions}.
\newblock Soft Matter. 2012;8(47):11933-8.

\bibitem{Lu_2012}
Lu PJ, Giavazzi F, Angelini TE, Zaccarelli E, Jargstorff F, Schofield AB,
  et~al.
\newblock {Characterizing Concentrated, Multiply Scattering, and Actively
  Driven Fluorescent Systems with Confocal Differential Dynamic Microscopy}.
\newblock Phys Rev Lett. 2012;108(21):218103.

\bibitem{Bayles_2016}
Bayles AV, Squires TM, Helgeson ME.
\newblock {Dark-field differential dynamic microscopy}.
\newblock Soft Matter. 2016;12(8):2440-52.

\bibitem{Martinez_2012}
Martinez VA, Besseling R, Croze OA, Tailleur J, Reufer M, Schwarz-Linek J,
  et~al.
\newblock {Differential Dynamic Microscopy: A High-Throughput Method for
  Characterizing the Motility of Microorganisms}.
\newblock Biophys J. 2012;103(8):1637-47.

\bibitem{Murphy_2012}
Murphy DB, Davidson MW.
\newblock {Fundamentals of Light Microscopy and Electronic Imaging}.
\newblock John Wiley {\&} Sons, Inc.; 2002.

\bibitem{Hoffman_1975}
Hoffman R, Gross L.
\newblock {Modulation Contrast Microscope}.
\newblock Appl Opt. 1975;14(5):1169.

\bibitem{Allen1969}
Allen R, David G.
\newblock {The Zeiss-Nomarski differential interference equipment for
  transmitted-light microscopy}.
\newblock Z Wiss Mikrosk. 1969;69(4):193-221.

\bibitem{Hamilton_2022}
Hamilton S, Regan D, Payne L, Langbein W, Borri P.
\newblock {Sizing individual dielectric nanoparticles with quantitative
  differential interference contrast microscopy}.
\newblock The Analyst. 2022;147(8):1567-80.

\bibitem{Reufer_2012}
Reufer M, Martinez VA, Schurtenberger P, Poon WCK.
\newblock {Differential Dynamic Microscopy for Anisotropic Colloidal Dynamics}.
\newblock Langmuir. 2012;28(10):4618-24.

\bibitem{HolmesAO87}
Holmes TJ, Levy WJ.
\newblock Signal-processing characteristics of
  differential-interference-contrast microscopy.
\newblock Appl Opt. 1987;26(18):3929.

\bibitem{Preza1999}
Preza C, Snyder DL, Conchello JA.
\newblock {Theoretical development and experimental evaluation of imaging
  models for differential-interference-contrast microscopy}.
\newblock J Opt Soc Am A. 1999;16(9):2185-99.

\bibitem{MunroOE05}
Munro PRT, Török P.
\newblock Vectorial, high numerical aperture study of Nomarski's differential
  interference contrast microscope.
\newblock Opt Express. 2005;13(18):6833.

\bibitem{Wang_2022}
Wang Y, Pope I, Brennan-Craddock H, Poole E, Langbein W, Borri P, et~al.
\newblock {A primary effect of palmitic acid on mouse oocytes is the disruption
  of the structure of the endoplasmic reticulum}.
\newblock Reproduction. 2022;163(1):45-56.

\bibitem{Regan_2019}
Regan K, Wulstein D, Rasmussen H, McGorty R, Robertson-Anderson RM.
\newblock {Bridging the spatiotemporal scales of macromolecular transport in
  crowded biomimetic systems}.
\newblock Soft Matter. 2019;15(6):1200-9.

\bibitem{Giavazzi_2014}
Giavazzi F, Cerbino R.
\newblock {Digital Fourier microscopy for soft matter dynamics}.
\newblock J Opt. 2014;16(8):083001.

\bibitem{Germain_2016}
Germain D, Leocmach M, Gibaud T.
\newblock {Differential dynamic microscopy to characterize Brownian motion and
  bacteria motility}.
\newblock Am Jo Phys. 2016;84(3):202-10.

\bibitem{Hansen2013}
Hansen JP, McDonald IR.
\newblock {Theory of Simple Liquids With Applications to Soft Matter}.
\newblock Elsevier Science \& Technology Books; 2013.

\bibitem{Berne2000}
Berne BJ, Pecora R.
\newblock {Dynamic light scattering: with applications to chemistry, biology,
  and physics}.
\newblock Courier Corporation; 2000.

\bibitem{Poon_2015}
Poon WCK.
\newblock In: Terentjev EM, Weitz DA, editors. {Colloidal Suspensions}. Oxford
  University Press; 2015. .

\bibitem{Shribak2017}
Shribak M, Larkin KG, Biggs D.
\newblock {Mapping optical path length and image enhancement using quantitative
  orientation-independent differential interference contrast microscopy}.
\newblock J Biomed Opt. 2017;22(1):016006.

\bibitem{Regan2019}
Regan D, Williams J, Borri P, Langbein W.
\newblock {Lipid bilayer thickness measured by quantitative DIC reveals phase
  transitions and effects of substrate hydrophilicity}.
\newblock Langmuir. 2019;35(43):13805-14.

\bibitem{Abramowitz1964}
Abramowitz M, Stegun IA.
\newblock {Handbook of mathematical functions with formulas, graphs, and
  mathematical tables}. vol.~55.
\newblock US Government printing office; 1964.

\bibitem{Mecholsky_2021}
Mecholsky NA, Akhbarifar S, Lutze W, Brandys M, Pegg IL.
\newblock {Dataset of Bessel function $J_n$ maxima and minima to 600 orders and
  10000 extrema}.
\newblock Data in Brief. 2021;39:107508.

\end{thebibliography}


\begin{thebibliography}{10}

\bibitem{Bayles_2016}
Bayles AV, Squires TM, Helgeson ME.
\newblock {Dark-field differential dynamic microscopy}.
\newblock Soft Matter. 2016;12(8):2440-52.

\bibitem{Giavazzi2009}
Giavazzi F, Brogioli D, Trappe V, Bellini T, Cerbino R.
\newblock {Scattering information obtained by optical microscopy: Differential
  dynamic microscopy and beyond}.
\newblock Phys Rev E. 2009;80(3):031403.

\bibitem{Giavazzi_2014}
Giavazzi F, Cerbino R.
\newblock {Digital Fourier microscopy for soft matter dynamics}.
\newblock J Opt. 2014;16(8):083001.

\bibitem{Preza1999}
Preza C, Snyder DL, Conchello JA.
\newblock {Theoretical development and experimental evaluation of imaging
  models for differential-interference-contrast microscopy}.
\newblock J Opt Soc Am A. 1999;16(9):2185-99.

\bibitem{Mehta2008}
Mehta SB, Sheppard CJ.
\newblock {Partially coherent image formation in differential interference
  contrast (DIC) microscope}.
\newblock Opt Express. 2008;16(24):19462-79.

\bibitem{Li2009}
Li K, Kanade T.
\newblock {Nonnegative mixed-norm preconditioning for microscopy image
  segmentation}.
\newblock In: {International Conference on Information Processing in Medical
  Imaging}; 2009. p. 362-73.

\bibitem{Koos2016}
Koos K, Moln{\'a}r J, Kelemen L, Tam{\'a}s G, Horvath P.
\newblock {DIC image reconstruction using an energy minimization framework to
  visualize optical path length distribution}.
\newblock Scientific reports. 2016;6(1):1-9.

\bibitem{VanHove1954}
Van~Hove L.
\newblock {Correlations in space and time and Born approximation scattering in
  systems of interacting particles}.
\newblock Phys Rev. 1954;95(1):249.

\bibitem{Hansen2013}
Hansen JP, McDonald IR.
\newblock {Theory of Simple Liquids With Applications to Soft Matter}.
\newblock Elsevier Science \& Technology Books; 2013.

\bibitem{Vineyard1958}
Vineyard GH.
\newblock {Scattering of slow neutrons by a liquid}.
\newblock Phys Rev. 1958;110(5):999.

\end{thebibliography}
\end{document}

\maketitle
In this work, we follow the simulation methodolgy outlined in Ref. \cite{Bayles_2016}. Simulated images depict 150 particles undergoing Brownian motion, with initial locations being uniformly randomly distributed throughout a square domain with length $600 \, \mu$m, partitioned into a $600 \times 600$ pixel grid such that the length of a pixel is $ 1\,  \mu \text{m}$. At each time step, each particle is given a displacement in both the horizontal and vertical direction, drawn from a normal distribution $\mathcal{N}(0, 2 D \Delta t)$ with diffusion coefficient $D=0.5 \mu \text{m}^2/\text{s}$ and time step $\Delta t = 0.125$s. In this way we construct a set of particle locations $(x_n(t),y_n(t)), $ $ n \in \{1,...,150\} $ for $4000$ time steps (hence, the total time is 500 s).

We assume that the Point Scattering Function, or PSF (See Section \ref{Section: ISF_Image}) applies to each particle as a Gaussian separately, 
\[X(\bm{r},t) = X(i,j,t) = \sum_{n=1}^{150}\alpha e^{-\frac{\left((i-x_n(t))^2 + (j-y_n(t))^2 \right)}{2\sigma^2}},\]
where $\alpha = 50$ (a.u) describes the magnitude of the phase shift and $\sigma=3$ pixels$^{-2}$ prescribes the effective beam width \cite{Bayles_2016}. The phase contributed by each particle is additive. 

\begin{figure}[H]
	\centering
	\includegraphics[scale=.7]{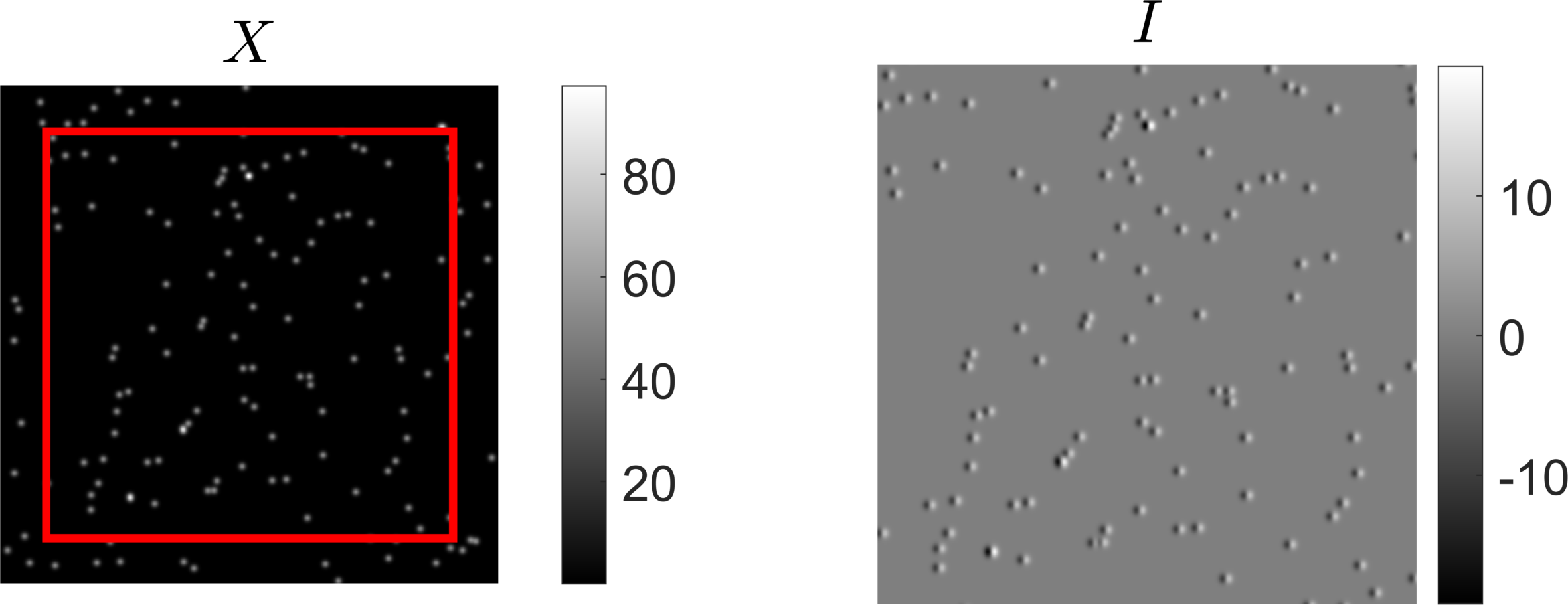}
	\caption{The first frame of the simulated phase image $X$ is found in the red square at the centre of the imaging domain, left. The corresponding DIC image, $I$, is generated by subtracting $X(\bm{r},\Delta t)$ from its shifted partner $Y(\bm{r},\Delta t)$ which is the image formed by shifting the red square right by one pixel.}
	\label{Fig: XandI}
\end{figure}

We assume periodic boundary conditions, so that particles exiting the domain over one boundary re-enter the domain through the opposite boundary; this ensures conservation of the particle number to preserve ergodicity. The image stack $X(\bm{r},\Delta t)$ is taken as a 490-pixel square window in the centre of the domain, visualised as the red square in Figure \ref{Fig: XandI}, such that the boundaries are not included in the images. Given our timescale $t=500$s and $D=0.5\mu \text{m}^2/\text{s}$, the Mean Square Displacement (MSD) for these simulations in two dimensions is 
\[\text{MSD} = 4 D t = 1000 \mu \text{m}^2,\]
implying the expected displacement is much smaller than the image window size. As such, it is unlikely that particles will travel a sufficiently large distance to exit the imaging window, cross the boundary and re-enter on the opposite side; hence it is not necessary to account for this type of motion, which simplifies the problem.

We also generate the image stack $Y(\bm{r},\Delta t)$ by shifting the imaging window right by a shear distance of 1 pixel, or $\bs = (1,0)$ $ \mu \text{m}$. The effective DIC image stack is defined as $I(\bm{r},\Delta t)=Y(\bm{r},\Delta t)-X(\bm{r},\Delta t)$. We display the first frame of the stacks $X(\bm{r},\Delta t)$ and $I(\bm{r},\Delta t)$ in Figure \ref{Fig: XandI}.

\section{The relationship between images and probability density functions in Fourier space}
\label{Section: ISF_Image}
 When imaging multiple objects, it is convenient to consider their density in space, rather than their specific position. We define spatial density by $c(\bm{r},\Delta t)$, which may either be a continuous scalar field (for example, concentration of a fluid) or, for $N$ discrete particles,
\begin{align}
	c(\bm{r},t) = \sum_{i=1}^{N} \delta(\bm{r}-\bm{r}_n(t)), \label{eq: particle_density}
\end{align}

where $\delta(\bm{r})$ is the Dirac delta function. We consider the relationship between the time-dependent image intensity distribution $I(\bm{r},t)$, and the time-dependent 3D sample density $c(\bm{r},z,t)$, to be Linear Space-Invariant (LSI), \cite{Giavazzi2009}, i.e
\begin{align}
	I(\bm{r},t) = i_0 + \int dz \int d^2 \bm{r}' K(\bm{r}-\bm{r}',z) c(\bm{r},z,t), \label{equation: LSI}
\end{align}

where $i_0$ is some approximately constant background illumination, $z$ is the location of the object in the axis orthogonal to the imaging plane, and $K(\bm{r}-\bm{r}',z)$ is known as the \textit{Point Spread Function} (PSF) \cite{Giavazzi_2014}, which describes how an object is visualised by microscopy. Whilst there is an established PSF for DIC images \cite{Preza1999,Mehta2008, Li2009}, it has also been shown that a linear approximation of the PSF coincides with the form of Equation \eqref{equation: LSI} \cite{Koos2016}. 
%An image stack being LSI is a sufficient, but not necessary, condition such that the same information can be extracted from a time series stack of microscope images, as from static or Dynamic Light Scattering (DLS) experiments \cite{Giavazzi_2014}. This idea forms a crucial basis for the following explanation of the DDM technique.

%Whilst there is an established PSF for DIC images \cite{Preza1999,Mehta2008, Li2009}, it has also been shown that a linear approximation of the PSF coincides with the form of Equation \eqref{equation: LSI} \cite{Koos2016}.

We describe the way in which particles move via the van Hove function, defined by \cite{VanHove1954}
\begin{align}
	G(\Delta \bm{r}, \Delta t) = N^{-1} \left \langle \sum_{i,j =1}^{N} \int \delta(\Delta \bm{r} +  \bm{r}-\bm{r}_j(t)) \delta( \bm{r}- \bm{r}_i(0)) d \bm{r}\right \rangle, \label{eq: VanHove_def}
\end{align}

where $\langle \cdot \rangle$ denotes the ensemble average or temporal average, which are equivalent for ergodic systems \cite{Hansen2013}. The van Hove function may be physically interpreted as the number of particles $j$ located within the unit volume $\Delta \bm{r}$ at time $t$, given a particle $i$ is located at the origin at time $t=0$ \cite{Vineyard1958, Hansen2013}. We can easily separate $G(\Delta x, \Delta t)$ into the sum of the `self' and `distinct' parts, $G_s(\Delta \bm{r},\Delta t)$ and $G_d(\Delta \bm{r},\Delta t)$, for when the $i=j$ or $i \neq j$ respectively, as follows \cite{VanHove1954}:
\begin{align*}
	G(\Delta \bm{r}, t) &= G_s(\Delta \bm{r}, t)+G_d(\Delta \bm{r}, t),\\
	G_s(\Delta \bm{r}, \Delta t) &= N^{-1} \left \langle \sum_{i =1}^{N} \int \delta(\Delta \bm{r} +  \bm{r}-\bm{r}_i(t)) \delta( \bm{r}- \bm{r}_i(0)) d  \bm{r}\right \rangle,\\
	G_d(\Delta \bm{r}, \Delta t) &= N^{-1} \left \langle \sum_{j\neq l=1}^{N} \int \delta(\Delta \bm{r} +  \bm{r}-\bm{r}_j(t)) \delta( \bm{r}- \bm{r}_i(0)) d \bm{r}\right \rangle,
\end{align*}
When particle trajectories are statistically independent, such that displacement of a particle is not affected by the position or displacement of any other particle, $G_d(\Delta \bm{r},\Delta t) = 0$, and we need only consider the self-correlating term $G_s$. By inspection, we see that $G_s$ effectively describes the probability that any given particle undergoes a displacement $\Delta \bm{r}$ during the time period $\Delta t$.

The relationship between particle density, $c(\bm{r},t)$, and the van Hove function, $G(\Delta \bm{r}, \Delta t)$, is given by \cite{Hansen2013}

\begin{align}
	G(\Delta \bm{r},\Delta t) &=  \frac{1}{N} \left \langle \sum_{l,j =1}^{N} \int \delta(\Delta \bm{r}+ \bm{r} - \bm{r}_l(t)) \delta( \bm{r}-\Delta \bm{r}_j(0)) d \bm{r}'\right \rangle \nonumber \\ 
	&=  \frac{1}{N} \left \langle \int c(\Delta \bm{r}+\bm{r},t)c(\bm{r},0) d \bm{r} \right \rangle \nonumber \\
	&= \frac{1}{\rho}\left \langle c(\Delta \bm{r},\Delta t) c(0,0) \right \rangle, \label{VanHove to C}
\end{align}
where $\rho = N/V$ is the particle density describing the expected number of particles found in a given volume $V$. Denoting the Fourier transform by $\tilde{\cdot}$ and the complex conjugate by $\cdot^*$, we define the Intermediate Scattering Function (ISF), as \cite{Hansen2013, Giavazzi_2014}
\begin{align} 
	f(\bm{q},\Delta t)	&= \frac{1}{N}\left \langle \tilde{c}(\bm{q},\Delta t) \tilde{c}^*(\bm{q},0) \right \rangle = \tilde{G}(q,\Delta t), \label{ParticleDensityRelationship}
\end{align}
where the second equality is derived by applying Parsevals theorem to Equation \eqref{VanHove to C}. Hence, the ISF allows us to relate particle density and the van Hove function in the Fourier domain. This is useful when images can be expressed by Equation \eqref{equation: LSI}, since we can define
\begin{align}
	S(\bm{q},\Delta t)=\frac{\langle \tilde{I}^*(\bm{q},t+\Delta t)  \tilde{I}(\bm{q},t) \rangle_t}{ \langle |\tilde{I}(\bm{q},t)|^2 \rangle_t} = \frac{\int dq_z |\tilde{K}(\bm{q},q_z)|^2 \langle \tilde{c}(\bm{q},q_z, t+\Delta t)^*  \tilde{c}(\bm{q},q_z,t) \rangle_t}{\int dq_z |\tilde{K}(\bm{q},q_z)|^2 \langle |\tilde{c}(\bm{q},q_z,0)|^2 \rangle_t}. \label{eq: DDM_PSF}
\end{align}
In systems where dynamics can be well approximated in two dimensions, for example in forms of microscopy where the image generated is a projection to the 2D plane, or where motion in the $z$ direction is negligible, the $q_z$ dependence can be neglected, and \eqref{eq: DDM_PSF} becomes
\begin{align}
	S(\bm{q},\Delta t)=\frac{\langle \tilde{c}(\bm{q}, t+\Delta t)^*  \tilde{c}(\bm{q},t) \rangle_t}{|\tilde{c}(\bm{q},0)|^2}. \label{eq: VHFT}
\end{align}
From Equation \eqref{ParticleDensityRelationship}, we can clearly see that
\begin{align} 
	S(\bm{q},\Delta t) = \frac{f(\bm{q},\Delta t)}{f(\bm{q},0)}.
\end{align}
Therefore, from the image stack $I(\bm{r},\Delta t)$, we can extract information about the Fourier transform of the van Hove function. We may further simplify this relationship in the case where particle trajectories are independent: In this case, $f(\bm{q},0) = 1$ \cite{Hansen2013}, and we deduce

\[	S(\bm{q},\Delta t) = \tilde{G}(\bm{q},\Delta t).\]

Therefore, DDM works by numerically determing $S(\bm{q},\Delta t)$, and then comparing to some known analytic form of $\tilde{G}(\vec{q},\Delta t)$ in order to extract specific parameters governing the rate of movement within the movie.

\bibliographystyle{vancouver}
%\bibliography{biblio}

\bibliography{biblio}